\documentclass[]{sig-alternate-10pt}

\usepackage{tikz}
\usepackage{amsmath}

\usepackage{blindtext}
\usepackage[normalsize,TABBOTCAP]{subfigure}
\usepackage{footmisc}

\usepackage{filecontents}
\usepackage{algpseudocode}
\usepackage{xspace}

\usepackage{verbatim}
\usepackage{color}
\usepackage{verbatim}
\usepackage{url}
\usepackage{graphicx}
\graphicspath{
	{./figure}
}
\usepackage{balance}
\usepackage{hyperref}
\usepackage{color}
\usepackage{comment}
\usepackage{makecell}
\usepackage{caption}

\usepackage{amsmath}
\usepackage{algorithm}
\usepackage{array}
\usepackage[toc,page]{appendix}
\usepackage{xspace}
\usepackage{xfrac}
\usepackage{amssymb}
\usepackage{mathrsfs}
\usepackage{wrapfig}
\usepackage[blocks]{authblk}

\def\ShowComment{1} % to show comment set 1, otherwise set 0.
\if\ShowComment1
\newcommand{\songmin}[1]{\textcolor{red}{SM: #1}}
\fi

\if\ShowComment1
\newcommand{\kangmin}[1]{\textcolor{green}{KM: #1}}
\fi

\if\ShowComment1
\newcommand{\hankyeol}[1]{\textcolor{blue}{HK: #1}}
\fi

\newcommand{\ours}{\texttt{Hawkeye}\xspace}

\usepackage{amsmath}

\begin{document}
    % \title{Hectometer-range Large-scale mmWave Backscatter Localization with Subcentimeter Accuracy}
    % \title{Hectometer-range Large-scale Subcentimeter \\ 3D Localization using mmWave Backscatter}
    \title{Hawkeye: Hectometer-range Subcentimeter Localization for Large-scale mmWave Backscatter}

\author[ ]{\Large{\textsf{Kang Min Bae$^{\dagger}$, Hankyeol Moon$^{\dagger}$, Sung-Min Sohn$^{\S}$, Song Min Kim$^{\P}$
}}}
\setlength{\affilsep}{0.1em}
\affil[ ]{\normalsize{Korea Advanced Institute of Science and Technology (KAIST) \\ $^{\S}$Arizona State University}}
\affil[ ]{\normalsize{\{bkm2259, moonkyul1, songmin\}@kaist.ac.kr, smsohn@asu.edu}}
% \renewcommand*{\thefootnote}{\arabic{footnote}}

% \authornote{Song Min Kim is the corresponding author (songmin@kaist.ac.kr).}
% \author{Kang Min Bae$^{\dagger}$, Hankyeol Moon$^{\dagger}$, Sung-Min Sohn$^{*}$, Song Min Kim$^{\S}$ \\
% Korea Advanced Institute of Science and Technology (KAIST)\\ $^{\S}$Arizona State University\\
% $^{\dagger}$Co-primary Student Authors}
% \{bkm2259, moonkyul1, songmin\}@kaist.ac.kr, smsohn@asu.edu}
% \affil{
% %     % \institution{
%          Korea Advanced Institute of Science and Technology (KAIST) \\ $^{\S}$Arizona State University
% %     % }
% }
% \affil[ ]{\{{bkm2259, moonkyul1, songmin}@kaist.ac.kr, smsohn@asu.edu}
% \email{{bkm2259, moonkyul1, songmin}@kaist.ac.kr, smsohn@asu.edu}
% \affiliation{%
%   \institution{Korea Advanced Institute of Science and Technology (KAIST) \\ $^{\S}$Arizona State University}
% }

\maketitle
\begingroup\def\thefootnote{$\dagger$}\footnotetext{Co-primary Student Authors.}\endgroup
\begingroup\def\thefootnote{$\P$}\footnotetext{Song Min Kim is the corresponding author.}\endgroup

	%A category including the fourth, optional field follows...
	%\category{D.2.8}{Software Engineering}{Metrics}[complexity measures, performance measures]
	
	%\terms{Delphi theory}
	
	%\keywords{ACM proceedings, \LaTeX, text tagging}
  		\begin{abstract}
	
% Pervasive interaction with the Internet of Things calls for precise and long-range localization at scale. 

Accurate localization of a large number of objects over a wide area is one of the keys to the pervasive interaction with the Internet of Things. This paper presents \ours, a new mmWave backscatter that, for the first time, offers over (i) hundred-scale simultaneous 3D localization at (ii) subcentimeter accuracy for over an (iii) hectometer distance. \ours generally applies to indoors and outdoors as well as under mobility. \ours tag's Van Atta Array design with retro-reflectivity in both elevation and azimuth planes offers 3D localization and effectively suppresses the multipath. \ours localization algorithm is a lightweight signal processing compatible with the commodity FMCW radar. It uniquely leverages the interplay between the tag signal and clutter, and leverages the spetral leakage for fine-grained positioning. Prototype evaluations in corridor, lecture room, and soccer field reveal $6.7~mm$ median accuracy at $160~m$ range, and simultaneously localizes 100 tags in only $33.2~ms$. \ours is reliable under temperature change with significant oscillator frequency offset.

	\end{abstract}
  	\begin{table*}
% \hspace{-0.2cm}
\centering
\small
\begin{tabular}{||c | c c c c c ||} 
 \hline
 Systems & \thead{Accuracy @ 5~m} & \thead{Range} & \thead{Bandwidth} & \thead{Simultaneous Localization} & \thead{Fixed Trajectory} \\
 \hline\hline
 \textbf{\ours} & \textbf{2.5~mm} & \textbf{180~m} & \textbf{250~MHz} & \textbf{100 Tags (1024 in theory)} & \textbf{No} \\ 
 \hline
 Millimetro~\cite{soltanaghaei2021millimetro} & 78~mm & 180~m & 250~MHz & 6 Tags (106 in theory) & No \\ 
 \hline
 RFind~\cite{ma2017minding} & 3.4~mm & 6~m & 220~MHz & No & No \\
 \hline
TurboTrack~\cite{luo20193d} & 5.1~mm & 10~m & 100~MHz & 2 Tags & No \\
 \hline
 Tagoram~\cite{yang2014tagoram} & 10~mm & 12~m & 6~MHz & No & Yes \\
 \hline
\end{tabular} \\

 % \vspace{-0.15in}
\caption{Comparison with the state-of-the-arts}\label{tab:comparison}
 \vspace{-0.2in}
\end{table*}

\section{Introduction}

%P1: State the problem that we are trying to solve, and stress the importance of the problem by discussing killer applications -- what would be the world like, with our technology?
Precise interaction with a large number of objects spread over a region has long been a vision for the IoT, where accurate localization is one of the essential features for an immersive experience. Backscatter possess great potential towards this goal, with the low-cost and ultra low-power tags that can be massively deployed over a large area with the minimum deployment cost and maintenance efforts. Localization of large-scale (e.g., hundreds to thousands) tags with subcentimeter accuracy installed over an area (e.g., hectometer-range) would offer benefits to a wide range of applications including asset tracking, inventory management, warehouse automation, smart factories, virtual/augmented reality, and structural health monitoring.

\begin{figure}[t!]
    \centering
    % \vspace{-4mm}
    \includegraphics[width=\columnwidth]{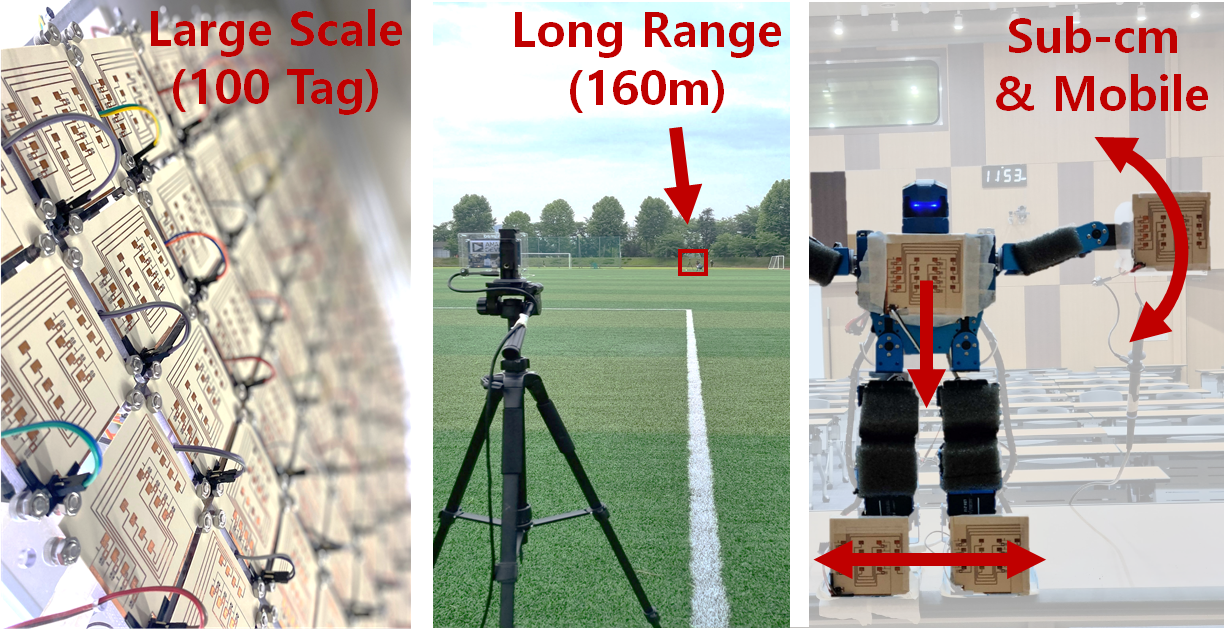}
    % \vspace{-7.5mm}
    \caption{\ours is tested under large-scale (left), over a long range (center), and on mobile objects (right).}
    % \textbf{\caption{The SNR boost and tag signal demodulation with and without HD-FMCW.}}
    \label{fig:front}
    \vspace{-3mm}
\end{figure}

%P2: Specific problem statement. Why should we work on this?
To this end, backscatter (including RFID) localization has been extensively studied in sub-6GHz bands. However, their accuracy, scalability, and range are fundamentally throttled by the hard bandwidth constraint (e.g., tens of KHz for UHF RFID). This limits their performance to tens of $cm$ accuracy~\cite{bouet2008rfid,chawla2013real,mao2007wireless,ni2003landmarc,zhou2009rfid}, restricts the deployment scenarios by requiring fixed movement trajectories~\cite{miesen2011holographic,shangguan2016design,yang2014tagoram, zhang2021siloc}, or requires deploying dense reference tags with prior knowledge~\cite{bouet2008range, han2015twins, wang2013dude}. A recent line of research, RFind~\cite{ma2017minding} and TurboTrack~\cite{luo20193d}, resolve the bandwidth issue by enabling RFID to emulate the wide bandwidth of 220 MHz that extends beyond the ISM band, to achieve subcentimeter accuracy. However, the range is bounded to several meters to remain compliant to the FCC regulations, limiting the usage to a room-scale and requiring a customized reader. The latest work of millimetro~\cite{soltanaghaei2021millimetro} takes a fundamental approach of exploring the rich, 250 MHz bandwidth in the 24 GHz mmWave band, by utilizing FMCW radar and backscatter tag to reach over $100~m$ range. However, the median accuracy of millimetro is limited to $15~cm$, which is essentially the accuracy of 24 GHz FMCW radar.

% XXXXX \songmin{why?}. 
% proposes a mmWave backscatter that leverages the 250MHz bandwidth of the 24GHz ISM band. Specifically it adopts 24GHz FMCW radar as backscatter reader to reach over 100m range, but the accuracy is limited 15cm due to XXXXX \songmin{why?}. 

%P3: high-level description on our design, with highlights on the advantages. Representative results should come here.
This paper presents \ours (Figure~\ref{fig:front}), a mmWave backscatter localization with the empirical performance of (i) $6.7~mm$ median accuracy (ii) at $160~m$ range (@1D), (iii) simultaneously localizes 100 tags in only $33.2~ms$ (scales up to 1024 tags in theory), and (iv) uses an affordable commodity radar ($\sim$200 USD~\cite{distance2go}). \ours blends a new backscatter tag for efficient signal delivery and lightweight radar-side signal processing for accurate and rapid localization. \ours backscatter tag is a planar Van Atta Array (VAA) combined with a power-efficient low-loss FSK modulator using hybrid coupler. The tag retro-reflects in both azimuth ($90^\circ$ FoV) and elevation ($140^\circ$ FoV) to enable 3D localization and effectively suppresses multipath. The design is robust against oscillator frequency offset, with only 4~$mm$ localization error across low ($9.45^\circ C$), high ($38.43^\circ C$), and room ($23.7^\circ C$) temperatures. This is an essential property for practical subcentimeter localization under disparate deployment scenarios, using low-end tags. Furthermore, one-shot interrogation localizes up to 1024 tags (evaluated with 100) simultaneously, supporting large-scale rapid localization. Table~\ref{tab:comparison} summarizes the comparison to the state-of-the-art backscatter localization systems, showcasing that \ours is uniquely positioned to achieve high scalability, long range, and precision at the same time.

%P4: Design details. What did we actually do and why is it so smart?
% \vspace{-1.2mm}
\ours exploits the interplay between \ours back-scatter FSK signal and the chirp-based Frequency Modulated Continuous Wave (FMCW) radar to improve the localization performance by over $\times$60 over using the FMCW alone. \ours tag is tuned to demonstrate S11 of -10 dB throughout the entire 250 MHz bandwidth, where FSK modulation is performed by the combination of reflective network and low-loss $90^\circ$ hybrid coupler co-optimized for efficient VAA reflection. The use of the VAA, along with the severe signal attenuation of the mmWave, naturally suppresses the multipath inteference. To the best of our knowledge, \ours tag is the first planar VAA mmWave backscatter design with modulation capability. The radar-side subcentimeter localization is designed as a lightweight post-processing on top of FMCW demodulation, without requiring any change to the commodity FMCW radar. Specifically, \ours localization algorithm is built on the recent technique of HD-FMCW~\cite{omniScatter} that isolates the tag FSK signal from the clutter noise. The key insight of \ours is to leverage the relationship between the tag signal and the clutter, and the spectral leakage signature embedded in the tag signal, from which the precise location can be extracted. \ours supports single radar or multilateration positioning, where 1D-3D localization was evaluated throughout indoors (corridors and lecture rooms), outdoors (soccer field), NLOS, and varying temperatures to demonstrate practicality.

% to benefit from its $>$50 dB SNR gain and the resilience to clutter noise -- briefly speaking, the disparate periods of the FSK tag signal and the interrogation signal (i.e., series of chirps) places the tag signal in the frequency bins different from those containing the clutter noise (which maintains the interrogation signal period), thereby disentangling the tag signal from the clutter noise. \ours essentially converts this SNR gain to fine-grained localization accuracy, by selectively eliminating the effect of FSK to yield clean, uncontaminated signal only containing the range, followed by the frequency resolution boosting. 

%P5: Enlist contributions and state the paper structure.
\ours is an accurate, long-range, and large-scale localization for mmWave backscatter, essentially enabling tracking many objects spread over an area, ranging from everyday spaces like homes and offices, to industrial sectors such as inventories and warehouses. \ours is kept economic with low-cost tags and compatibility to affordable commodity radar. To sum up, we believe \ours takes a solid step towards bringing pervasive tag deployment and localization to practice. Our contribution is three-fold:

\vspace{-2mm}
\begin{itemize}

\item We design \ours, a mmWave backscatter-based subcentimeter localization that works over hecto-meter-range and simultaneously localizes over a thousand tags.

\vspace{-0.9mm}
\item To the best of our knowledge, we design the first planar VAA mmWave backscatter with modulation capability.

\vspace{-0.9mm}
\item We prototype \ours tags on Rogers RO4003C substrate for antenna with planar VAA structure, VXCO-based control board on the PCB. Hundred tags were produced for large-scale simultaneous 3D localization. Readers were implemented on commodity 24GHz radars~\cite{tinyrad, distance2go}. 

\vspace{-0.9mm}
\item We will release \ours's source code and HFSS tag design file upon acceptance, for facilitating community's future works. 

% Extensive testbed evaluations spanning indoor, outdoor, blockages, and temperature were performed to demonstrate practical impact.

\end{itemize}
\vspace{-0.7mm}

  	\section{Background}
This section provides the technical background for \ours, followed by the design overview.

% \subsection{Preliminaries}

\vspace{2mm} \noindent {\bf FMCW Radar.} An FMCW radar leverages chirp, whose frequency linearly increases with time. Objects in the radar's vicinity reflect the transmitted chirp, which returns to the radar with a round-trip propagation delay. FMCW radar mixes the transmitted chirp with the received chirp (with propagation delay) to produce an IF signal. The range is measured by performing FFT on the IF signal, where each reflected object is represented as a signal with range frequency $f_r$ proportional to the propagation delay.
% \kangmin{range frequency should be explained.}

% \vspace{1mm} \noindent {\bf Backscatter.} Backscatter achieves ultra-low power communication at only tens of $\mu W$. Essentially, the low power consumption is achieved by omitting the power-hungry passband signal generation~\cite{jung2020gateway}, by modulating the wireless signal on reflection. A backscatter tag switches between different impedance to control the reflection coefficient~\cite{reflection_coef}, which alters the amplitude, frequency, and phase of wireless signal~\cite{OFDMbackscatter}.

% \vspace{0.1in} \noindent {\bf Backscatter Localization.}
% \kangmin{?}

\vspace{0.1in} \noindent \textbf{Planar Van Atta Array.}
Planar Van Atta array (VAA) passively reflects back the signal to the direction of arrival, achieving retro-reflectivity in both azimuth and elevation planes. It is a simple antenna array structure where centrosymmetric pairs of the antenna in 2D plane are interconnected by transmission lines (TLs) with a length difference equal to $\lambda_{g}$ (i.e., guided wavelength, which is the wavelength of EM wave in the dielectric).
\setlength{\columnsep}{10pt}
\begin{wrapfigure}{l}{0.15\textwidth}
\centering
\vspace{-0.4cm}
% \subfigure[Linear VAA]{\includegraphics[width=0.55\columnwidth]{figures/linear_VA.png}}
% \hspace{3mm}
\includegraphics[width=0.15\textwidth]{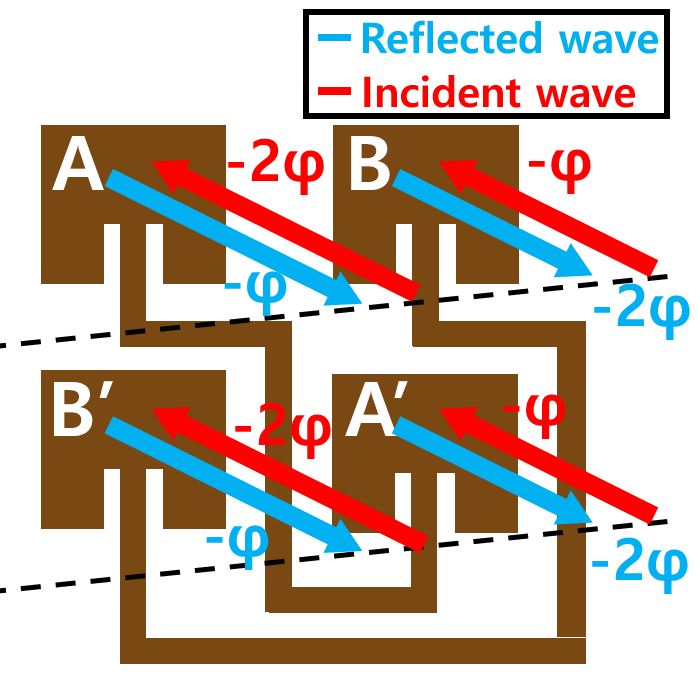}
\vspace{-6mm}	
\caption{Planar Van Atta array}\label{fig:va_back}
\vspace{-0.5cm}
\end{wrapfigure}
The centrosymmetrically interconnected antenna pair flips the incident signal's phase sequence, which directs the signal to the source.
The phase induced by the TLs does not affect the radiation direction, because it is applied equally to all lines.
As an illustrative example in Figure~\ref{fig:va_back}, consider a 2$\times$2 planar VAA, where the signal comes in the azimuth plane with the phase sequence of [$-2\varphi$, $-\varphi$] at both [A,B] and [B',A'].
After the propagation in TL, the phase sequence is inverted and produces a reflected signal with a phase sequence of [$-\varphi$, $-2\varphi$] at both [A,B] and [B',A'].
This makes the reflected wave back to the incident angle achieving retro-reflectivity in the azimuth plane.
Retro-reflectivity in the elevation plane is also achieved in the same manner, by flipping the phase sequence at [A,B'] and [B, A'].

\begin{figure}[h!]
\centering
\vspace{-2mm}
\includegraphics[width=0.55\columnwidth]{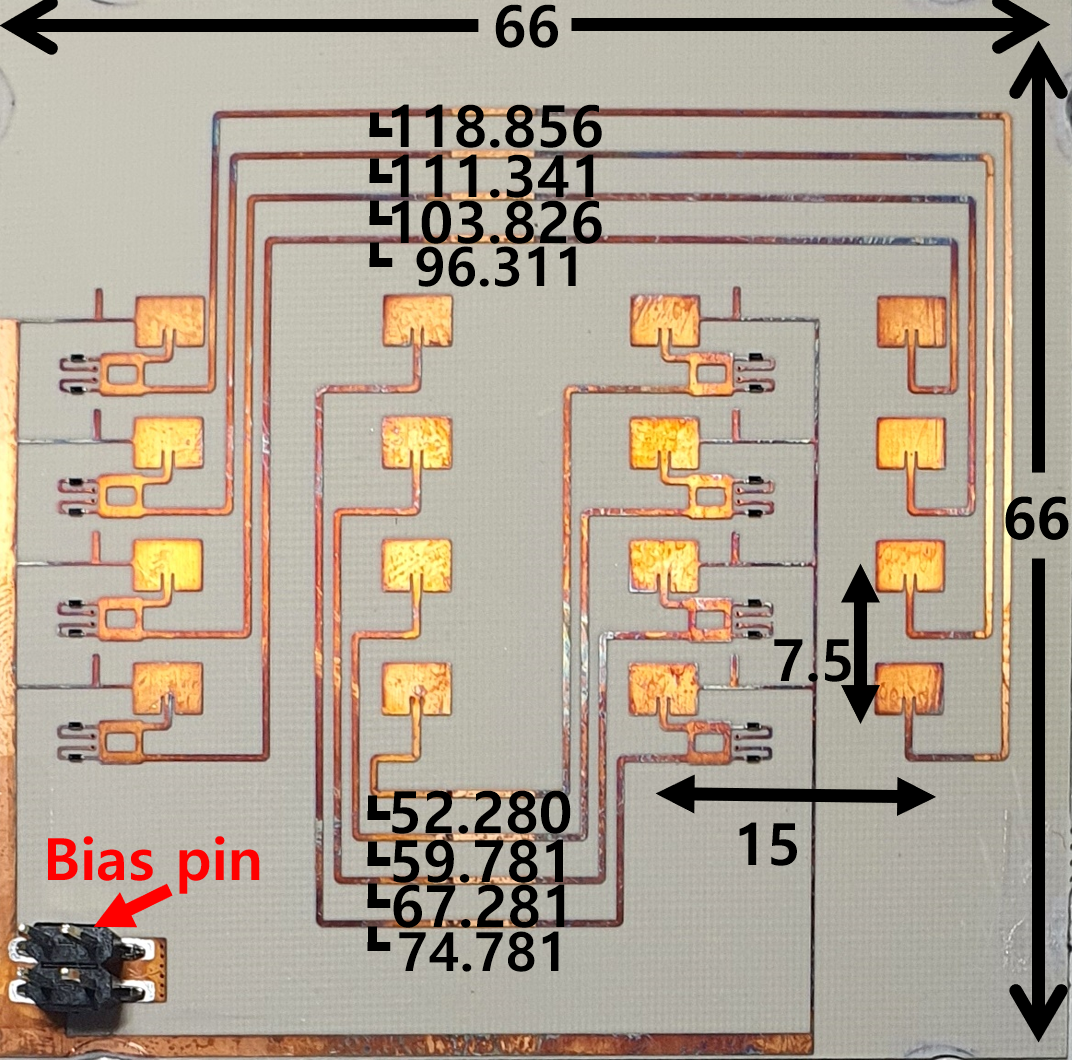}
% \subfigure[]{\includegraphics[width=0.55\columnwidth]{figures/tag_partX_alone.png}}
\vspace{-2mm}
\caption{\ours tag with key geometrical parameters optimized for the 24 GHz band. 
Centrosymmetric antenna pairs are interconnected by TLs.
% Retro-reflectivity is achieved by the interconnection of centrosymmetric inset-fed patch antennas, inverting the phase between incident and reflected signals.
All dimensions are in $mm$.}
\label{fig:planarVA}
\vspace{-0.4cm}
\end{figure}

% \begin{figure}[h!]
% % \centering
% \subfigure[]{\includegraphics[width=0.425\columnwidth]{figures/tag_whole.png}}
% \subfigure[]{\includegraphics[width=0.558\columnwidth]{figures/tag_part.png}}

% \vspace{-3mm}
% \caption{\ours planar VAA tag with key geometrical parameters optimized for the 24 GHz band. (a)  Retro-reflectivity is achieved by the interconnection of centrosymmetric inset-fed patch antennas, inverting the phase between incident and reflected signals. (b) Magnified view. Inset-fed patch antenna and TLs are optimized to match the impedance. Hybrid coupler operates with the PIN diodes for FSK modulation. All dimensions are in $mm$.}
% \label{fig:planarVA}
% \vspace{-0.2cm}
% \end{figure}

\begin{figure}[h!] 
% \centering
\vspace{-4mm}
\subfigure[]{\includegraphics[width=0.48\columnwidth]{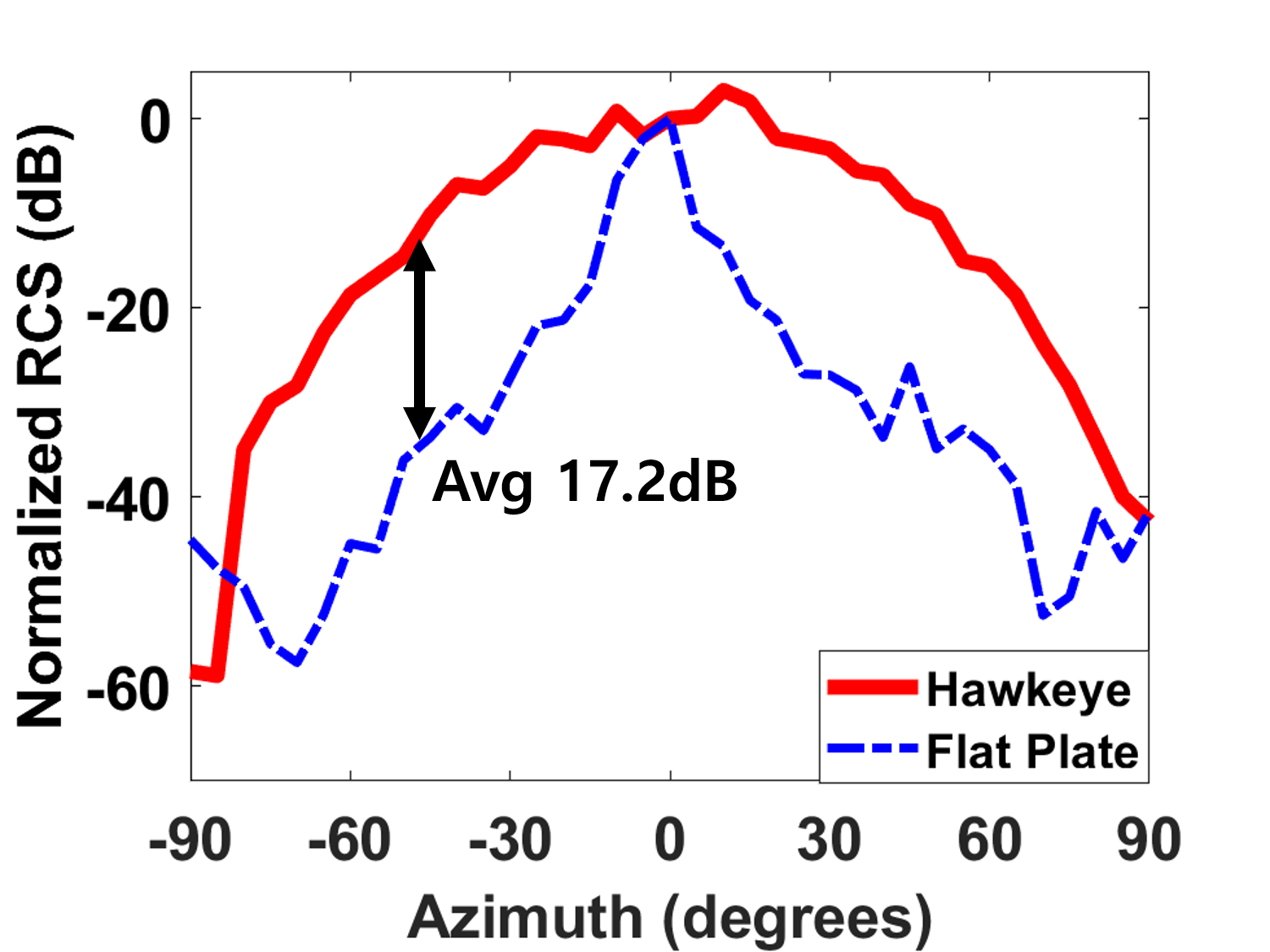}}
\subfigure[]{\includegraphics[width=0.48\columnwidth]{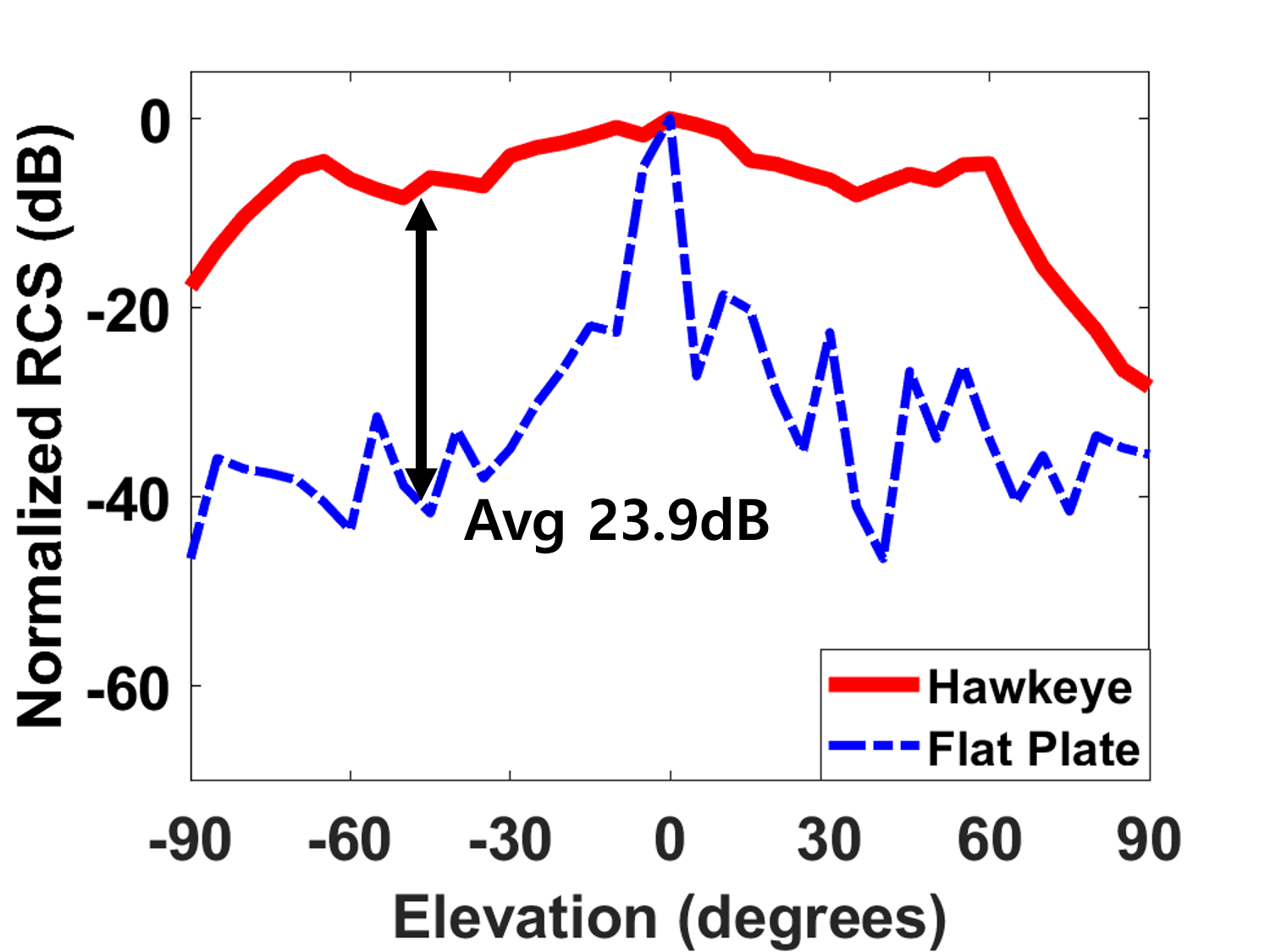}}
\vspace{-5mm}	
\caption{Comparison of measured normalized monostatic RCS of \ours tag and equal-sized flat plate with the same substrate (a) in azimuth plane and (b) in elevation plane. 
In each cases, we achieve over $17.2$ dB and $23.9$ dB beamforming gain over the flat plate.}\label{fig:retro_graph}
\vspace{-0.3cm}
\end{figure}

% \begin{figure}[h!]
% \centering
% \includegraphics[width=\columnwidth]{figures/inclination_ver2.png}
% \vspace{-5mm}
% \caption{Comparison of maximum beam inclination/USD per number of patch antenna. When M=4, it gives the marginal gain per USD, which is shown as the the elbow point at graph.}
% \label{fig:inclination}
% \vspace{-0.4cm}
% \end{figure}

\section{Hawkeye Tag Design}\label{sec:tag}

Figure~\ref{fig:planarVA} presents \ours tag prototype, which uniq-uely blends a 4$\times$4 planar Van Atta array with FSK modulation which serves as a basis to long-range subcentimeter 3D localization with effective multipath suppression. In the following, we present detailed structure and design choices. 

\vspace{-2mm}
\subsection{Retro-reflectivity via Planar VAA}\label{sec:vanatta}
\vspace{-1mm}

As depicted in Figure~\ref{fig:retro_graph}, \ours VAA achieves retro-reflectivity with an average beamforming gain of $17.2$ dB with FoV (10dB beamwidth) of $90^\circ$ and $140^\circ$ in azimuth and elevation, respectively. 
% This retro-reflectivity spans 250 MHz in the 24 GHz band, ensuring compatibility with the commodity FMCW radars with 250 MHz bandwidth. 
To meet the VAA condition, centrosymmetric antenna pairs are interconnected via TLs, whose lengths are based on the guided wavelength of $\lambda_{g}=7.56~mm$, derived from the 24.125 GHz (i.e., the center frequency of 24GHz band). Specifically, the length differences of TLs are multiples of $\lambda_g$. To limit the phase misalignment within the 250MHz bandwidth, the difference between the maximum and the minimum TL lengths is capped to $9\lambda_g$. This translates to the maximum phase misalignment of $16.9^\circ$, or equivalently, beamforming power loss of only $5\times10^{-2}$dB.

% $\lambda_{g}=7.56~mm$ for minimum phase misalignment. 
% The phase misalignment grows as the interrogation signal (i.e., the FMCW chirp) frequency deviates from the center frequency, towards the boundary of the band at the edge of chirp. The frequency offset is integrated over the TL length differences, incurring the phase misalignment. \ours tag caps the difference between the maximum and the minimum TL lengths to $9\lambda_g$. The maximum TL difference of $9\lambda_g$, collectively with maximum frequency offset of 125 MHz, translates to maximum phase misalignment of $16.9^\circ$, or beamforming power loss of only $0.11$dB.

% In order to meet the VAA condition of phase inversion between the antenna pairs, our antenna array consists of 8 centrosymmetric antenna pairs whose TLs differ by the guided wavelength, $\lambda_g$.
%In other words, precise phase inversion throughout the 250 MHz bandwidth is achieved for optimal retro-reflectivity. 

\begin{figure}[h!]
\vspace{-0.3cm}
% \centering
\subfigure[S11]{\includegraphics[width=0.46\columnwidth]{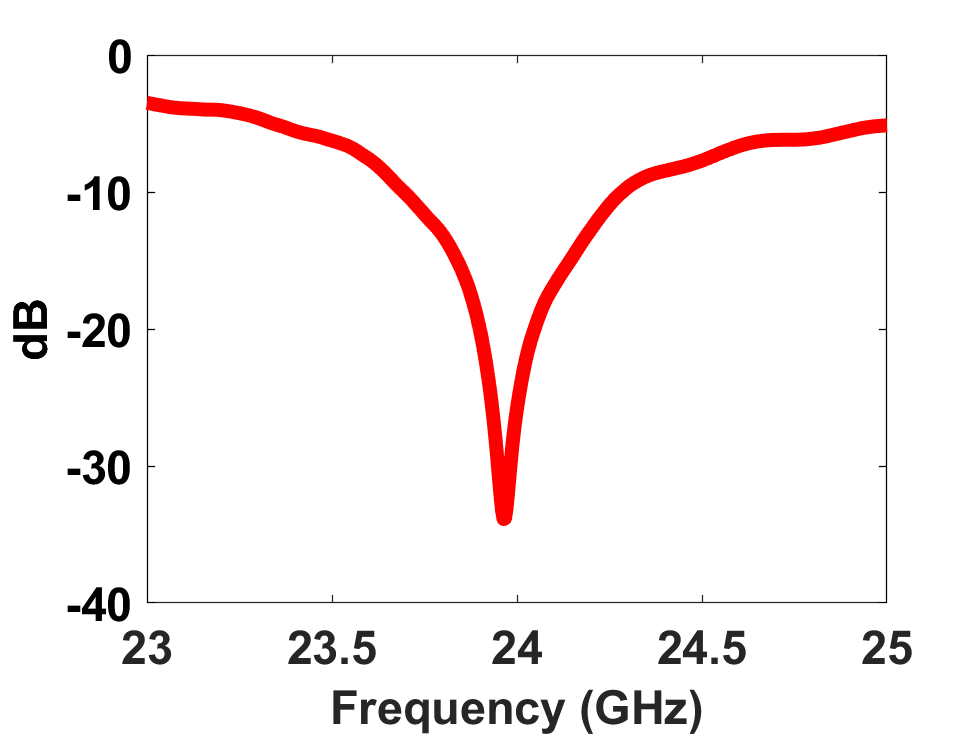}}
\hspace{2mm}
\subfigure[Smith chart]{\includegraphics[width=0.38\columnwidth]{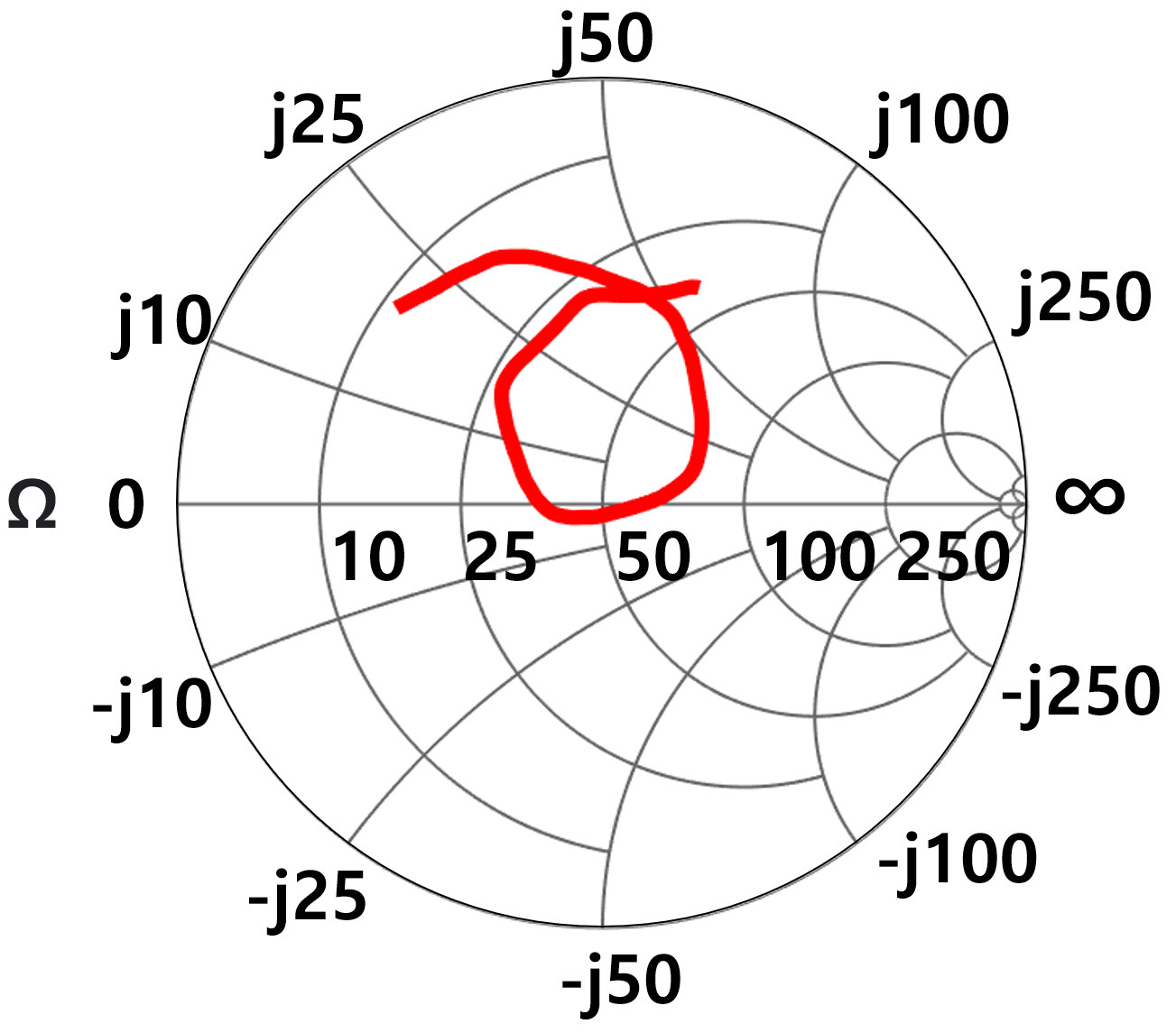}}
% \subfigure[]{\includegraphics[width=0.45\columnwidth]{figures/onoff_phase.png}}
\vspace{-3mm}	
\caption{
% Measured antenna performance of \ours tag, where (a) demonstrates our optimization results in S11 of $-10~dB$ throughout the 24 GHz band, and (b) shows the corresponding impedance matching result represented in Smith chart.
Measured antenna performance of \ours tag, with (a) S11 and (b) Smith chart.
}\label{fig:simul_result}
\vspace{-0.2cm}
\end{figure}

\begin{figure}[h!]
\centering
\vspace{-0.1cm}
\subfigure[]{\includegraphics[width=0.43\columnwidth]{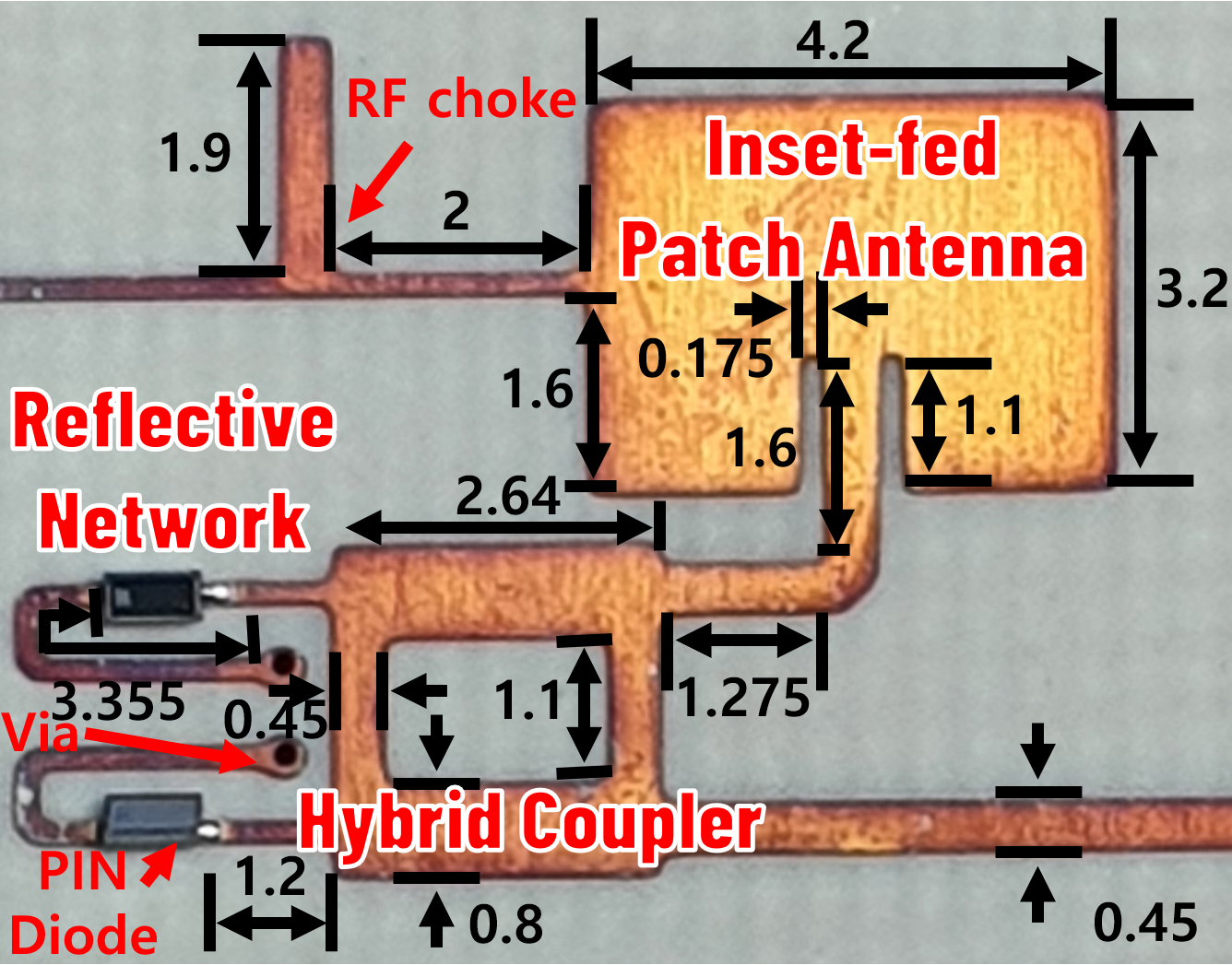}}
\subfigure[]{\includegraphics[width=0.535\columnwidth]{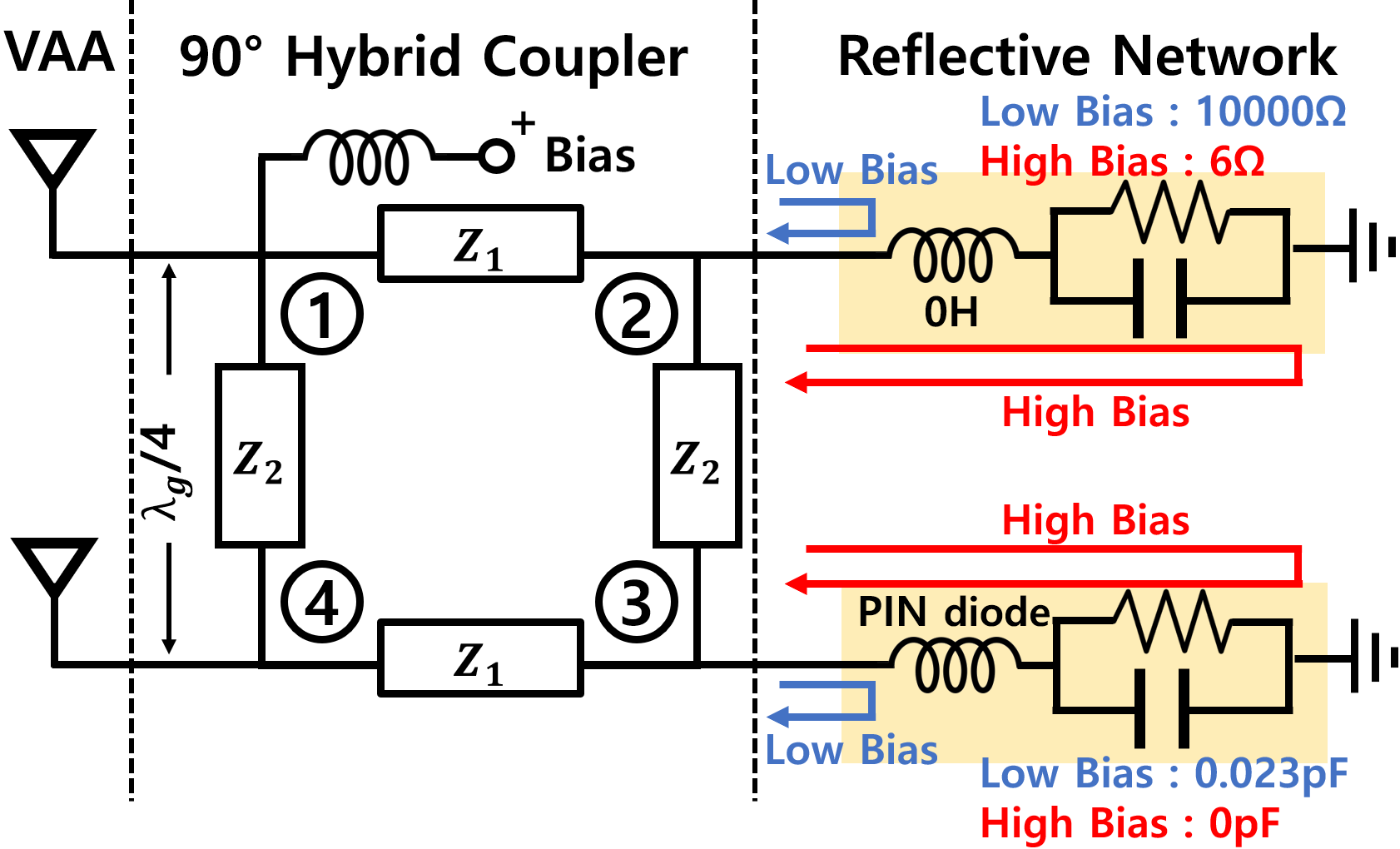}}
\vspace{1mm}
\vspace{-3mm}	
\caption{(a) Magnified view of \ours tag with key geometrical parameters. 
% Inset-fed patch antenna and TLs are optimized to match the impedance. 
% Hybrid coupler operates with the PIN diodes for FSK modulation. 
All dimensions are in $mm$. 
(b) Equivalent circuit of one pair of antenna, equipped with a $90^\circ$ hybrid coupler and a reflective network. 
% The ground symbol represents the via hole connected to the ground plate.
}\label{fig:phase_shifter}
\vspace{-0.3cm}
\end{figure}

\ours tag adopts inset-fed rectangular microstrip patch antenna with the patch, edge notch depth (i.e., the antenna feeding point~\cite{matin2010design}), and TL dimensions optimized to 50 $\Omega$ impedance matching at 24GHz, through HFSS parametric sweep. Figure~\ref{fig:simul_result} demonstrates the antenna performance measured via VNA; Figure~\ref{fig:simul_result}(a) depicts the antenna response (S11) of $-10$ dB throughout the 24 GHz band, and Figure~\ref{fig:simul_result}(b) shows the corresponding impedance matching results. The parameters are shown in Figure~\ref{fig:phase_shifter}(a) on Rogers RO4003C substrate (dielectric constant $\epsilon_{r}=3.55$, dissipation factor tan$\delta=2.7\times10^{-3}$), where the top layer holds \ours tag circuit while bottom layer is ground plate.
% For 50 $\Omega$ impedance matching, \ours tag adopts inset-fed rectangular microstrip patch antenna with an edge notch with tuned depth (i.e., the antenna feeding point~\cite{matin2010design}).
% \hankyeol{, which has the advantage of smaller size relative to its alternatives~\cite{cadence_microstrip}.} 
% \begin{figure}[h!]
% \centering
% \vspace{-0.1cm}
% \includegraphics[width=\columnwidth]{figures/inclination_v3.png}
% \vspace{-5mm}
% \caption{Comparison of maximum beam inclination/antenna element per number of patch antenna. When M=4, it gives the marginal gain per antenna element, which is shown as the the elbow point at graph.}
% \label{fig:inclination}
% \vspace{-0.3cm}
% \end{figure}
Lastly, 4$\times$4 antenna elements are chosen to balance between the FoV of the retro-reflectivity and the tag size -- i.e., the maximum beam inclination\footnote{Max beam inclination for antenna array is $90^\circ -47.83^\circ\times{(\lambda/Md)}^{1/2}$~\cite{934915_beamdeflection}, where $M$, $\lambda$, and $d$ are \# of antenna elements, wavelength, and antenna spacing, respectively.} increased beyond 4$\times$4 becomes marginal, relative to the exponentially increasing size.

\vspace{-2mm}
\subsection{Low-loss Modulator for Planar VAA}\label{sec:modulator}
% \hankyeol{structure description, why 1 -1, Result}
An efficient modulator directly affects the SNR of the tag signal, which determines the tag's detection robustness.
Given \ours operating throughout the 250 MHz bandwidth in the 24 GHz, this section discusses how it effectively performs FSK modulation without affecting the retro-reflectivity.
% Extensive care must be given to the FSK modulator design for operation in the 24 GHz mmWave, as its short wavelength (i.e., 12.5 mm) makes the signal extremely sensitive to even a slight error in the circuit.
% There are two major difficulties in designing a mmWave system: (i) high attenuation property in both air and substrate, and (ii) sensitive phase change corresponding to the circuit structure due to its short wavelength. Thus, our design of the modulator aims to (i) minimize the signal loss from the tag, and (ii) get the exact phase shifts by optimizing the dimension of each component, with consideration of the coupling effect.
Figure~\ref{fig:phase_shifter}(b) presents the equivalent circuit representation of an antenna pair, corresponding to Figure~\ref{fig:phase_shifter}(a). \ours tag consists of 8 such antenna pairs, each equipped with a $90^\circ$ hybrid coupler and a reflective network for FSK modulation via periodic $180^\circ$ phase shifts. 
% \hankyeol{Efficient FSK modulation via the 180$^\circ$ phase shift minimizes the power loss of the interrogation signal so as to reflect with the highest power. It retains the incident signal power, as opposed to toggling between absorption and reflection states -- losing half of the interrogation signal power (during absorption).}
As shown in Figure~\ref{fig:phase_shifter}(a), both the coupler and the reflective networks have microstrip architecture with a couple of PIN diodes. This structure enables \ours to maximize SNR by
% (i) maintaining the interrogation signal power with symmetric reflective networks 
% (i) maintaining the interrogation signal power with 180$^\circ$ phase shift
% (ii) low loss with the $90^\circ$ hybrid coupler, and 
(i) low-loss characteristic of the $90^\circ$ hybrid coupler in combination with symmetric reflective networks, and (ii) keeping the retro-reflectivity intact whilst FSK modulation via carefully designed bias and compensation of the coupling effects, avoiding leakage or distortion of the signal. Our design collectively brings low-loss FSK modulation, laying a solid foundation for hectometer-range support, in combination with the unique localization algorithm in  Section~\ref{sec:SuperResolution}.

% Phase Shifter vs. SPDT
%reflective network
% \hankyeol{del this paragraph}
\vspace{1mm}\noindent \textbf{Low-loss Phase Shifting using Hybrid Coupler.} 
\ours performs energy-efficient FSK via (i) $180^\circ$ phase shifts with (ii) Hybrid Coupler, minimizing the backscatter reflection loss at the \ours tag. As in Figure~\ref{fig:phase_shifter}(b), the phase is $180^\circ$ flipped depending on the length of the signal path at the reflection network, controlled by the ON (high bias) and OFF (low bias) states of the PIN diode. This retains the incident signal power for low-loss modulation (cf. many state-of-the-art backscatters modulate via SPDT switches, to toggle between absorption and reflection states~\cite{omniScatter, mazaheri2021mmtag, soltanaghaei2021millimetro} -- essentially sacrificing the half of the interrogation signal power (during absorption)). For retro-reflectivity, the modulated interrogation signal flows into the other side of the antenna pair. This is achieved with low-loss via a hybrid coupler and a reflective network in Figure~\ref{fig:phase_shifter}(b), performing as the impedance matched TL (details in the later part of the section). Alternative low-loss modulator is the switched line phase shifter that switches between two TLs with different lengths. However, this design requires $2\times$ diodes than the hybrid coupler~\cite{koul1991microwave} to induce higher power consumption and cost. While the switched line phase shifter has advantage in the formfactor, this advantage is insignificant in mmWave. 
% Further, \ours modulator still keeps a small size ($5.16~mm \times 2.7~mm$) to be contained in small tag, because it is built on mmWave frequency.

\begin{figure}[h!]
\centering
% \vspace{-0.5cm}
\includegraphics[width=\columnwidth]{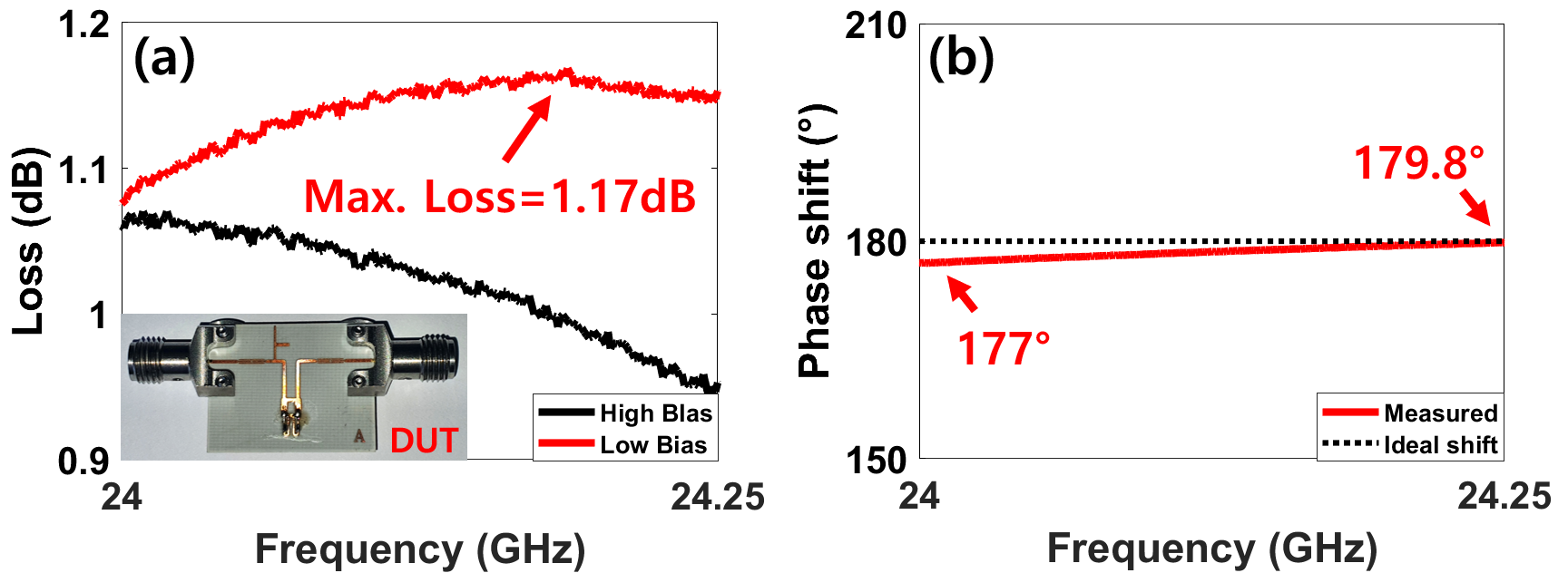}
\vspace{-5mm}	
\caption{
The measured response from the fabricated modulator unit throughout the 24GHz band. 
(a) Loss for low and high biases, (b) phase shift between low and high biases.}\label{fig:loss_graph}
\vspace{-0.3cm}
\end{figure}

To understand the operation of \ours phase shifter, let us refer to Figure~\ref{fig:phase_shifter}(b) where the incident interrogation signal is received at the upper antenna (port $\raisebox{.5pt}{\textcircled{\raisebox{-.9pt} {1}}}$) and flows into the lower antenna (port $\raisebox{.5pt}{\textcircled{\raisebox{-.9pt} {4}}}$). To maintain the retro-reflectivity alongside FSK modulation, the hybrid coupler performs low-loss signal transfer from the port $\raisebox{.5pt}{\textcircled{\raisebox{-.9pt} {1}}}$ to port $\raisebox{.5pt}{\textcircled{\raisebox{-.9pt} {4}}}$, with the reflective network in between. For this, the hybrid coupler acts as an impedance-matched TL with low insertion loss~\cite{koul1991microwave}, by the following operation:
The incident signal is divided into four paths by the coupler, each flowing into reflective networks and bouncing off back to the coupler -- i.e., $path_1$: $\raisebox{.5pt}{\textcircled{\raisebox{-.9pt} {1}}}\rightarrow\raisebox{.5pt}{\textcircled{\raisebox{-.9pt} {2}}}\rightarrow\raisebox{.5pt}{\textcircled{\raisebox{-.9pt} {1}}}$, $path_2$: $\raisebox{.5pt}{\textcircled{\raisebox{-.9pt} {1}}}\rightarrow\raisebox{.5pt}{\textcircled{\raisebox{-.9pt} {3}}}\rightarrow\raisebox{.5pt}{\textcircled{\raisebox{-.9pt} {1}}}$, $path_3$: $\raisebox{.5pt}{\textcircled{\raisebox{-.9pt} {1}}}\rightarrow\raisebox{.5pt}{\textcircled{\raisebox{-.9pt} {2}}}\rightarrow\raisebox{.5pt}{\textcircled{\raisebox{-.9pt} {4}}}$, and $path_4$: $\raisebox{.5pt}{\textcircled{\raisebox{-.9pt} {1}}}\rightarrow\raisebox{.5pt}{\textcircled{\raisebox{-.9pt} {3}}}\rightarrow\raisebox{.5pt}{\textcircled{\raisebox{-.9pt} {4}}}$.
% \hankyeol{check}
% \songmin{explain coupler with Z impedances to show the delivery betwee ports 1 and 3. explain without mentioning coupling effect}
% As aforementioned, the signal delivered diagonally (e.g., between port $\raisebox{.5pt}{\textcircled{\raisebox{-.9pt} {1}}}$ and $\raisebox{.5pt}{\textcircled{\raisebox{-.9pt} {3}}}$) gains additional $90^\circ$ phase compared to signal delivered horizontally (e.g., between port $\raisebox{.5pt}{\textcircled{\raisebox{-.9pt} {1}}}$ and port $\raisebox{.5pt}{\textcircled{\raisebox{-.9pt} {2}}}$) at $90^\circ$ hybrid couplers.
% ~\cite{koul1991microwave}
Hybrid coupler with four $\lambda_{g}/4$ TLs yields 180$^\circ$ phase difference between $path_1$ and $path_2$, and the same phase between $path_3$ and $path_4$~\cite{reed1956method}. By keeping the signal amplitudes in the four paths identical, $path_1$ and $path_2$ cancel each other at port $\raisebox{.5pt}{\textcircled{\raisebox{-.9pt} {1}}}$. Therefore, no signal is radiated on the upper antenna, thereby minimizing the reflection (i.e., $S_{11}$=0). On the other hand, $path_3$ and $path_4$ are constructively added at port $\raisebox{.5pt}{\textcircled{\raisebox{-.9pt} {4}}}$ for maximized radiation on the lower antenna (i.e., $S_{41}=1$). In summary, the incident signal is maximally delivered from the upper to the lower antenna, or equivalently, the insertion loss is minimized when the signal amplitudes in the four paths are the same. The path amplitude ratios are computed as $path_1/path_2=Z_{2}^2$ and $path_3/path_4=Z_2/Z_2$ when $Z_1 = Z_2/\sqrt{Z_{2}^{2}+1}$~\cite{reed1956method}. For the ratios of 1, we get $Z_{1} = 1/\sqrt{2}$ and $Z_{2} = 1$, that is, 35.3$\Omega$ and 50$\Omega$, respectively.
VNA measurement in Figure~\ref{fig:loss_graph}(a) shows the low-loss of under 1.17 dB, where the modulator was isolated from the tag (DUT in the figure) for precise measurement.

% \hankyeol{operation of overall component, + stub optimization through diode analyze}
\vspace{1mm}\noindent \textbf{Incorporating the Modulator and Planar VAA.} For precise $180^\circ$ phase shifts throughout the entire 250 MHz in the 24 GHz we leverage an equivalent circuit in Figure~\ref{fig:phase_shifter}(b) with the PIN diode MADP-000907-14020 represented in corresponding R, L, and C as per the datash-eet~\cite{madp_diode}. Also, to maintain VAA retro-reflectivity while combining with the FSK modulator, RF leakage through the bias line (driving the PIN diodes) should be avoided -- the leakage would distort the phase and corrupt the retro-reflectivity. \ours adopts $\lambda_{g}/4$ open stub (i.e., an open-ended microstrip line~\cite{hong2004microstrip}) as an RF choke to prevent RF leakage, which is represented as an inductor in the equivalent circuit. The parameter values are found through an extensive HFSS parameter sweep simulation, including the TL lengths and gaps between TLs to compensate for the coupling effect between the modulator and the TLs. The parameter values are shown in Figures~\ref{fig:planarVA} and~\ref{fig:phase_shifter}(a). As in Figure~\ref{fig:loss_graph}(b), \ours modulator yields an accurate FSK with the maximum error of 3$^\circ$ (i.e., [177$^\circ$, 179.8$^\circ$]), indicating only $3\times10^{-3}$dB power loss throughout the entire 250 MHz in the 24 GHz band. The RF signal was effectively isolated by greater than 30dB by the RF choke to achieve FSK modulated retro-reflectivity.

\begin{figure}[h!]
\vspace{-3mm}	
\centering
% \vspace{-0.5cm}
\includegraphics[width=\columnwidth]{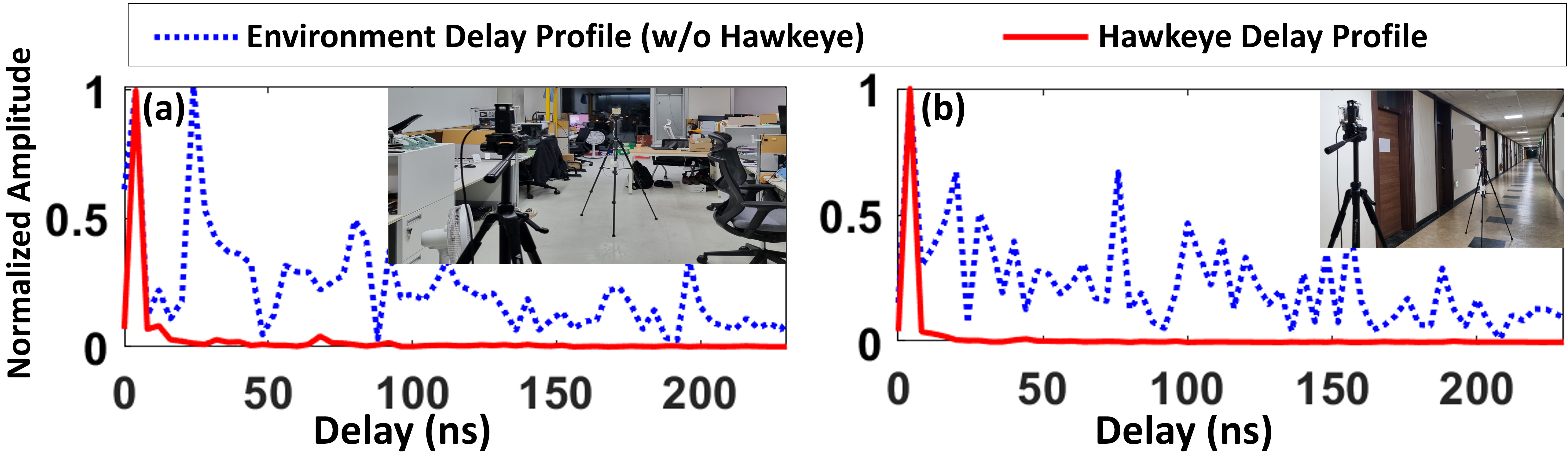}
\vspace{-6mm}	
\caption{Delay profiles in an (a) office, and a (b) hallway. Environment delay profile demonstrates the multipath-rich indoor scenario with the delay spanning over 230us. \ours effectively suppresses the multipath (i.e., delay) via retro-reflectivity. 
% \kangmin{Professor's comment: the hallway looks too large (multipath 안일어날것처럼생김), due to the resized photo. Please fix this!}
% For each experiment, the delay profile of the environment (without \ours) and \ours are captured,  to analyze how much multipath signal affects to \ours system.
}\label{fig:multipath}
\vspace{-0.2cm}
\end{figure}

% \vspace{-2mm}
\subsection{Indoor Multipath Suppression}\label{sec:multipath}

% Finally, \ours achieves planar VAA backscatter, with the FSK modulation with a maximum loss of only 1.175dB. 
% The retro-reflectivity of \ours tag gives about 11.89dB beamforming power gain relative to the omnidirectional tag, while keeping the gain over 17.2 dB over a flat plate. The enhanced signal power dramatically increases the detection range, enabling \ours to achieve 96~\% detection rate with 160 m tag-radar distance.
% especially at the mmWave band.
% The resulting efficient FSK modulator, in combination with the robust planar VAA, 
Retro-reflectivity of \ours effectively suppresses the multipath, enabling robust indoor localization. Figure~\ref{fig:multipath} presents the delay profile measured from two multipath-rich indoor settings of an office and a hallway. In both scenarios \ours significantly reduces the delay spread and limits the multipath signal power to be 20dB or more below the LOS signal. This is because, under retro-reflectivity, the received NLOS (i.e., multipaths) signals are strictly limited to the direction to which the signals are sent (instead of all directions without retro-reflectivity). Thus, higher the retro-reflectivity, less the delay spread.
% (ii) all NLOS signals (i.e., multipaths) undergoes two or more reflections and hence are significantly attenuated. For instance, 87\% signal power will be lost when a signal reflects twice on a drywall~\cite{zhao201328}.
% This stems from its retro-reflectivity, where the VAA reflects the multipath (reflected from the clutter) to the direction where it comes.
% It results that the multipath will be reflected from same clutter again, with significant reflection loss -- e.g., 87\% signal power will be lost when the multipath reflects from drywall twice~\cite{zhao201328}.
The multipath suppressed by \ours tag in Figure~\ref{fig:multipath} induces only $8.8~mm$ (office) and $2~mm$ (hallway) error in \ours localization (Section~\ref{sec:SuperResolution}), enabling subcentimeter indoor positioning as demonstrated in Section~\ref{sec:lDlocalize}.

\begin{figure}[h!]
\centering
\vspace{-4mm}
\subfigure[FMCW]{\includegraphics[width=0.49\columnwidth]{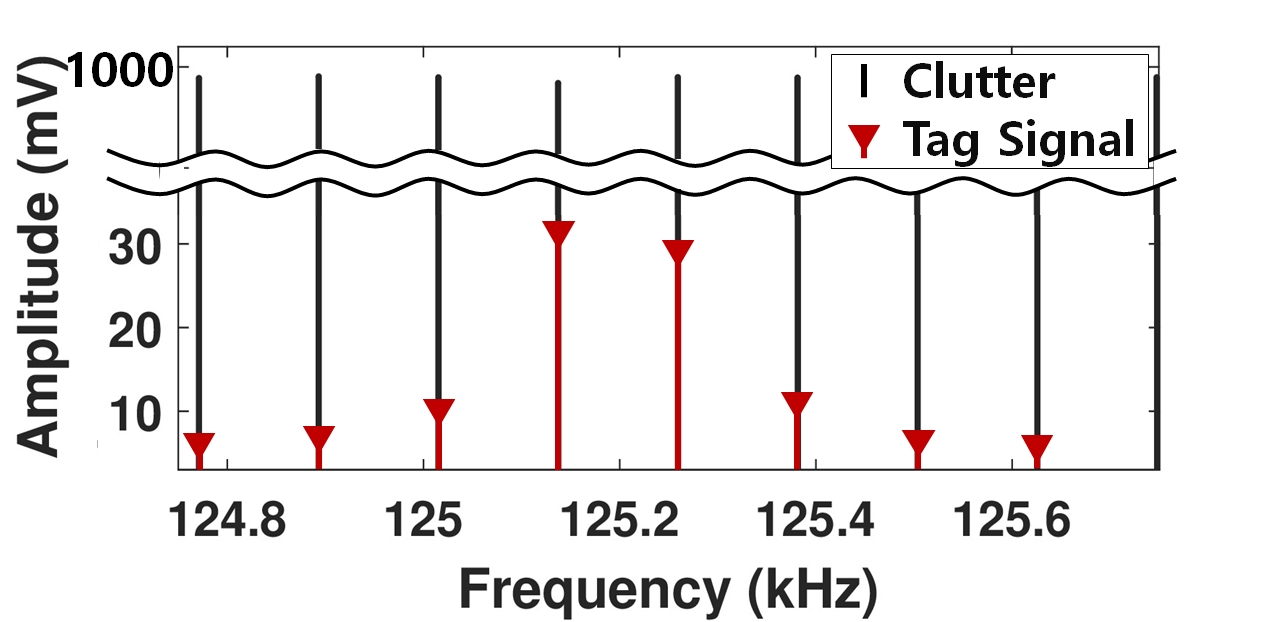}}
\subfigure[HD-FMCW~\cite{omniScatter}]{\includegraphics[width=0.49\columnwidth]{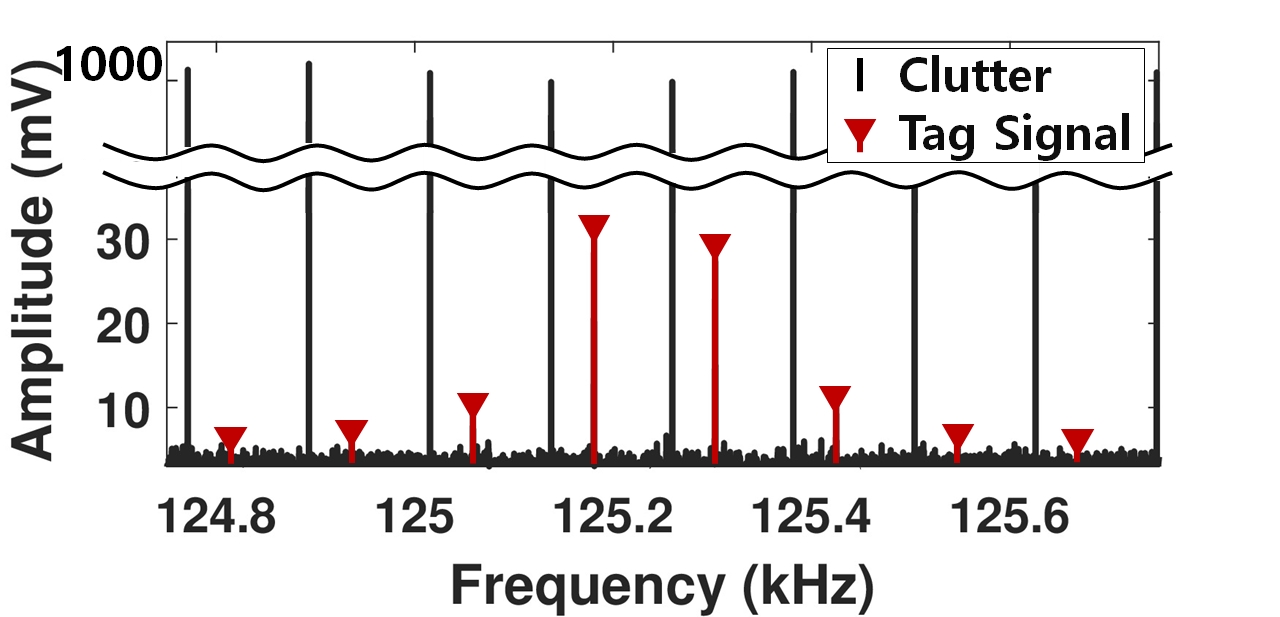}}
\vspace{-3mm}	
\caption{IF of FMCW and HD-FMCW with tag FSK at 40 Hz. HD-FMCW isolates tag signal from clutter.
% ($51.17~dB$ SNR gain).
}\label{fig:HD-FMCW}
\vspace{-5mm}
\end{figure}

% This drawing is for Section 4.
\begin{figure*}
    \centering
    % \vspace{-4mm}
    \includegraphics[width=2.1\columnwidth]{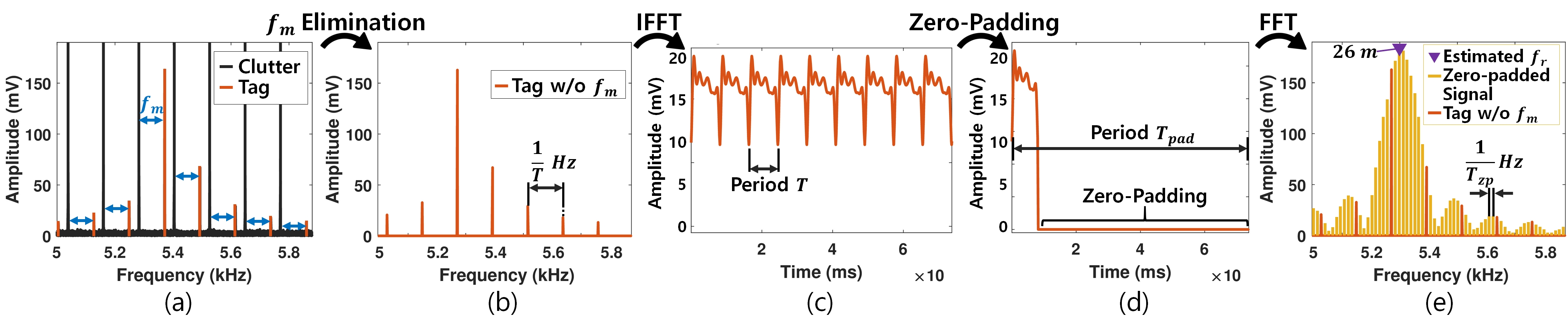}
    \vspace{-7.8mm}
    \caption{\ours one-shot localization process. The $f_m$ is determined in real time (a) to be detached from IF, where IFFT is applied to (b) the isolated clean signal with only $f_r$ (including the spectral leakage). At (c) time domain of the signal, (d) a subset with duration $T$ is zero-padded, where (e) FFT is performed to reveal the envelope sinc function. The frequency of the maximum peak amplitude is chosen as $f_r$.}
    % \textbf{\caption{The SNR boost and tag signal demodulation with and without HD-FMCW.}}
    \label{fig:SR}
    \vspace{-2mm}
\end{figure*}

\section{Hawkeye Localization}\label{sec:SuperResolution}
\vspace{4mm}

% \songmin{single interrogation, static, mobile, many}
In this section, we describe how \ours achieves subcentimeter localization at hectometer-range, using the commodity radar and \ours tag. Essentially, it is an extremely accurate ranging design that can be extended to 2D/3D localization with mutliple (multilateration) or single radar. At a high level, \ours leverages the spectral leakage signature of a tag signal to extract super-resolution range frequency.
Our design leverages a recent technique of HD-FMCW presented in OmniScatter~\cite{omniScatter}.
We first provide a brief primer on HD-FMCW, followed by the subcentimeter localization supporting mobile tags and large-scale simultaneous localization.

\vspace{3mm}\noindent \textbf{HD-FMCW Primer.} A recent technique of HD-FMCW, with a light add on signal processing on commodity FMCW radars, effectively isolates the FSK signal from the clutter noise in the frequency domain. Compared to the original FMCW which uses a single-chirp symbol, HD-FMCW leverages multiple (periodic) chirp symbol, $s(t)=c(t)\ast \sum_{n=1}^{N}\delta(t-nT)$, where $c(t)$ is a chirp with duration $T$, $N$ is the number of chirp repetitions, and $*$ is the convolution. This interrogation signal, when reflected from the clutter (i.e., clutter noise), simply becomes $s(t-\Delta t)$ where $\Delta t$ is the round trip propagation delay between the radar and the clutter. Since it is simply a time-shifted version of the interrogation signal, it maintains the period of $T$. This is therefore represented as peaks on the multiples of $\frac{1}{T}~\mathrm{Hz}$ frequency bins in the IF signal, where all other bins in between are left zero\footnote{From the Fourier's Theorem~\cite{FourierTheorem}, an arbitrary signal with period $T$ seconds in the time domain is represented as the multiples of $\frac{1}{T}~\mathrm{Hz}$ in the frequency domain.}. Note that, this applies to clutter noise from all sources -- i.e., all noises are concentrated on the same set of frequency bins, leaving other bins zero. On the contrary, the interrogation signal reflected off the tag (i.e., the tag signal) is not only time-shifted, but also modulated by the tag FSK. Specifically, the tag signal is

% \vspace{-0.1in}
\begin{equation}
% \begin{split}
\underbrace{s(t-\Delta t)}_{\substack{\text{Interrogation signal} \\ \text{(period $T$)}}}\cdot \underbrace{e^{j2\pi f_{m} t}}_{\substack{\text{FSK} \\ \text{(period $\frac{1}{f_m}$)}}}
% \end{split}
\label{eq:demodulation}
\end{equation}

\noindent where $f_m$ is the modulation frequency of the FSK. This yields a new period, other than $T$ (i.e., the least common multiple of $T$ and $\frac{1}{f_m}$). Therefore, in IF, tag signal and clutter noise are placed in different frequency bins -- isolating the tag signal from all environmental clutter noise. The separation of tag and clutter is demonstrated at Figure~\ref{fig:HD-FMCW}, where HD-FMCW (Figure~\ref{fig:HD-FMCW}(b)) isolates the tag signal unlike the original FMCW (Figure~\ref{fig:HD-FMCW}(a)).

\begin{figure}[h!]
\centering
\vspace{-0.12cm}
\includegraphics[width=0.85\columnwidth]{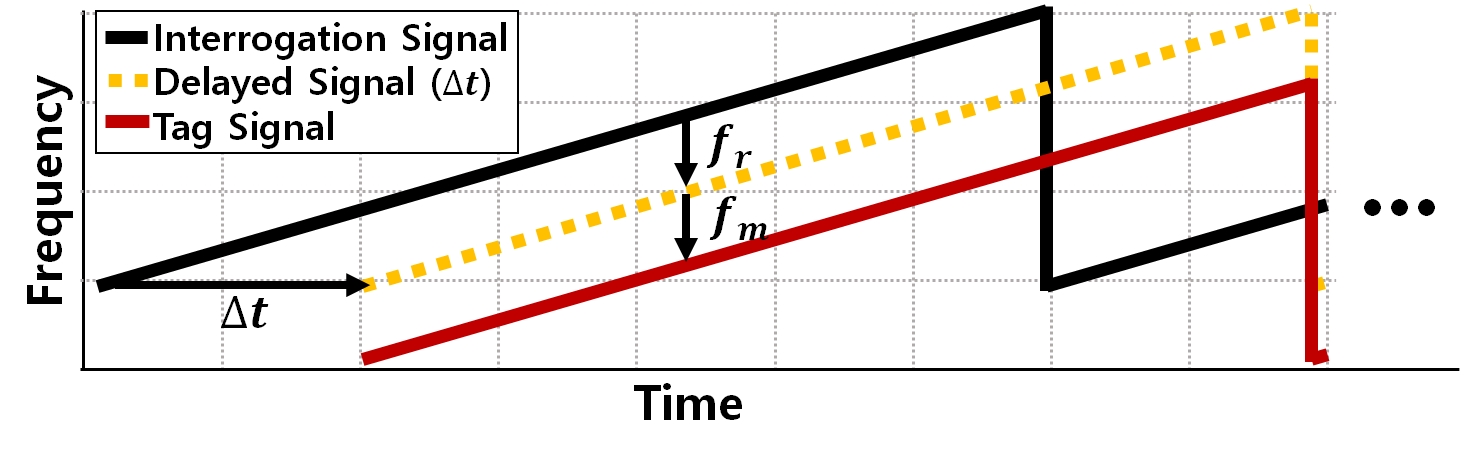}
\vspace{-5mm}	
\caption{The time-frequency representation of tag signal at HD-FMCW. The $\Delta t$ is the round trip propagation delay between radar and tag.}
\label{fig:fmcw_backscatter}
\vspace{-0.38cm}
\end{figure}

% \vspace{-1.5mm}
\subsection{Subcentimeter-accuracy Localization}\label{sec:subcm_loc}

% \ours provides localization accuracy of XX~mm up to distances of YY~meters.

\ours localization uniquely leverages the relationship between the isolated tag signal and the clutter noise in the HD-FMCW, and the spectral leakage signature embedded in the tag signal. Figure~\ref{fig:fmcw_backscatter} illustrates the reflected tag signal in relation to the interrogation signal, where the offset between the two signals is the combination of $f_r$ and $f_m$. The range frequency $f_r$ is from the tag-radar propagation delay $\Delta t$ whereas $f_m$ is the FSK modulation frequency. In other words, $f_r$ indicates the true position of the tag and therefore, accurate localization translates to the problem of finding $f_r$. This is achieved in two steps: (i) Removing the effect of $f_m$ from the relationship between the tag signal and clutter noise, and (ii) precise estimation of $f_r$ from the spectral leakage signature, which we discuss in the following.

% FSK modulation frequency $f_m$. \ours achieves subcentimeter localization by precisely deriving $f_r$. This is a two-step process -- (1) Detaching $f_r$from $f_m$ and applying the super-resolution frequency estimation leveraging the spectral leakage signature.
% % leveraging the spectral leakage signature.

% \vspace{-0.2mm}

% \ours's super-resolution algorithm stems from the observation that while the $f_r$ corresponds to the precise tag-radar distance, its representation is constrained by the frequency resolution of $1/T~Hz$ (i.e., the frequency resolution of FMCW radar) to limit the range accuracy. This phenomenon persists even with HD-FMCW, as the increased frequency-resolution of HD-FMCW only shifts the frequency modulated signals to the added frequency bins while maintaining the frequency quantization of $1/T~Hz$.
% Contrarily, our observation is that the exclusive tag signal isolation of HD-FMCW applies to the entire tag signal including the spectral leakage, which enables us to rigorously derive $f_r$ by exploiting its spectral leakage signature. As $f_r$ is a single-tone signal, its spectral leakage follows a sinc function precisely centered at $f_r$ due to the DFT property~\cite{oppenheim1975digital}. Thus, reconstructing the envelope sinc function allows \ours super-resolution to correctly derive the $f_r$ (i.e., the center of the reconstructed sinc) with boosted frequency resolution.

% \vspace{1mm} \noindent \textbf{Eliminating $f_m$.} 
\vspace{2mm} \noindent \textbf{Eliminating Tag Modulation $f_m$.} 
The first step to achieving accurate estimate of the range frequency $f_r$ is to eliminate the effect of $f_m$ in the IF signal. We note that, in practice, $f_m$ is not known in advance; It needs to be detected in real time due to the instability of the oscillator speed. For instance, a crystal oscillator can have $500~ppm$ variance under different environments~\cite{500ppm}. This indicates $25~\mathrm{Hz}$ error at $50~\mathrm{kHz}$, which can translate to a vast amount of over $24.6~cm$ localization error. \ours is inherently robust to the oscillator variance, as it detects the accurate $f_m$ on the fly, without any prior knowledge on the modulation speeds or the environment under which the tags are installed.

% This is in fact The $f_m$ is produced at tag whose oscillator may be unstable, and the inaccuracy may be critical in practice. For instance, a typical crystal oscillator of $150~ppm$ accuracy produces up to $15~Hz$ error at $100~kHz$, which translates to over $1.84~cm$ error at practical \ours settings. 

% \vspace{-0.5mm}
In order to precisely identify $f_m$, we begin from the fact that $f_r$ originates from the propagation delay between the radar and tag, or equivalently, $s(t-\Delta t)$ ( Eq.~\ref{eq:demodulation}) that has the period $T$. On the contrary, $f_m$ stems from the FSK signal of $e^{j2\pi f_{m} t}$ with period $\frac{1}{f_m}$. As a result, the IF signal is represented as peaks at frequencies with $\frac{1}{T}$ Hz interval (from $s(t-\Delta t)$), with offset of $f_m$ (from $e^{j2\pi f_{m} t}$) as depicted in Figure~\ref{fig:SR}(a). Thus, the $f_m$ can be precisely determined in real time, simply from the offset from $\frac{n}{T}~\mathrm{Hz}, (n\in\mathbb{N})$ frequency bins. We note that those frequency bins hold the clutter noise; Therefore, to remove $f_m$, the clutter noise are first nullified to zero, and then the tag signal is shifted to the nullified frequency bins. Figure~\ref{fig:SR}(b) depicts the resulting clean signal with only $f_r$, free of contamination from $f_m$ and clutter noise. This is a key technique for subcentimeter localization unaffected from the tag clock offset prevalent in practice.

% \vspace{1mm} \noindent \textbf{Super-resolution $f_r$.} 
\vspace{3mm} \noindent \textbf{Extracting Super-resolution Range Frequency $f_r$.} 
\ours subcentimeter localization is achieved by accurately identifying the $f_r$ (indicating tag-radar distance) with boosted frequency resolution. 
This deviates from the conventional HD-FMCW, where the $f_r$ is represented with frequency resolution of $\frac{1}{T}~\mathrm{Hz}$ -- i.e., identical to the original FMCW frequency resolution.
To do so, \ours exploits the spectral leakage in the discrete Fourier Transform (DFT), where the DFT of a signal with the period of $T$ and frequency of $f_r$ is represented as peaks at the multiples of $\frac{1}{T}~\mathrm{Hz}$ whose envelope follows the sinc function centered at $f_r$ -- i.e., $T\mathrm{sinc}(\pi T (f-f_r))$~\cite{oppenheim1975digital}. Therefore, accurately deriving $f_r$ becomes the problem of fine-grained identification of the envelope sinc function from which the center frequency can be pinpointed.
To achieve this, \ours zero-pads a subset with duration $T$ to increase the period to $T_{\mathrm{pad}}$ ($\gg T$) in the time domain, as depicted in Figures~\ref{fig:SR}(c),(d). Figure~\ref{fig:SR}(e) demonstrates the zero-padding result, where the peaks in the frequency domain are densified to precisely reveal the envelope sinc function. In our experiment we set $T_{\mathrm{pad}}=128T$ to keep the computation overhead low while achieving the subcentimeter localization accuracy. Given the dense peaks, the center frequency $f_r$ is simply found as the frequency with the maximum peak amplitude.

An extensive experiment reveals \ours median range error of $2.5~mm$, over $\times$60 improvement compared to the original FMCW with $15~cm$ median error.
% This conforms to the Cramér–Rao bound~\cite{Steven_Estimation}, where $50~dB$ SNR gain (from HD-FMCW) indicates two orders of magnitude accuracy improvement at maximum.
With the super-resolution $f_r$ acquisition mechanism, \ours achieves subcentimeter localization up to $160~m$ outdoors, and $80~m$ indoors. We note that entire \ours localization algorithm has the computation complexity of $O(N\log N)$ (for FFT/IFFT) where $N$ is the number of samples -- retaining the complexity of the original FMCW that mandatorily runs FFT.

\begin{figure}[h!]
\centering
\vspace{-0.1cm}
\includegraphics[width=0.8\columnwidth]{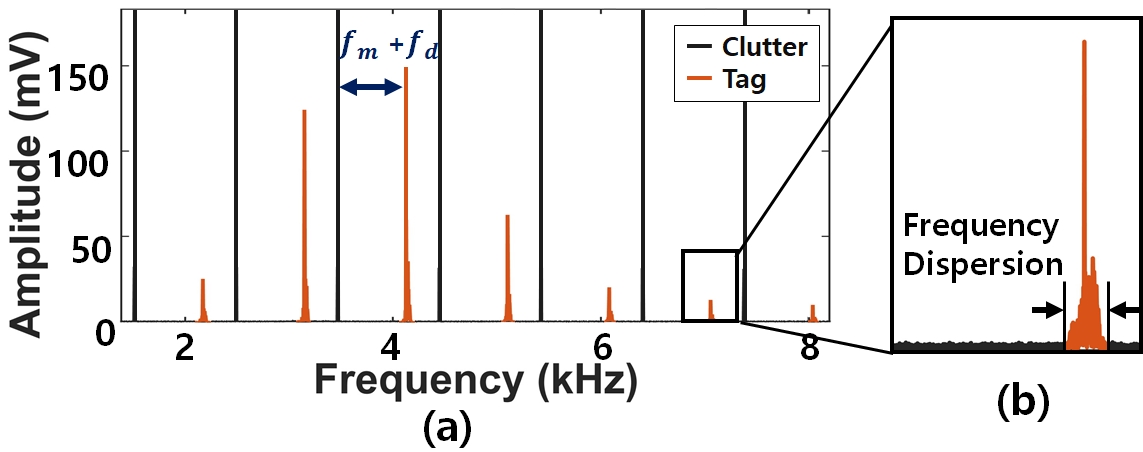}
\vspace{-2.5mm}	
\caption{IF signal of a mobile tag. (a) The tag signal has offset of $f_m+f_d$ from clutter noise, where (b) varying $f_r$ incurs frequency dispersion.}
\label{fig:fspread}
\vspace{-0.38cm}
\end{figure}

\subsection{Mobile Tags}\label{sec:tracking}

On the contrary to the static tags, a mobile tag induces Doppler frequency $f_d$ and time-varying range frequency $f_r(t)$. Figure~\ref{fig:fspread}(a) illustrates the IF of the mobile tag ($\leftrightarrow$ Figure~\ref{fig:SR}(a) for static tag), where $f_d$ is added on top of $f_m$. The $f_d$ is effortlessly removed together with $f_m$ by simply following the $f_m$ elimination mechanism discussed in the previous section. On the other hand, time-varying range frequency, $f_r(t)$, causes frequency dispersion of the peak as shown in Figure~\ref{fig:fspread}(b). \ours tracks $f_r(t)$ with subcentimeter accuracy, through fine-grained temporal analysis. For mobile localization, we begin by distinguishing the moving tags from the static ones via frequency dispersion, proportional to the tag velocity\footnote{This is known as the dispersion factor~\cite{dispersion} in radar context}. For subcentimeter localization, we define mobile tags as those with $>1~cm$ movement within a symbol, revealed to be the frequency dispersion of $\geq 1.4~Hz$ according to our empirical study. Then, the movement is tracked by the following.

\vspace{1mm}\noindent{\textbf{Extracting Time-varying Range Frequency $f_r(t)$.}}
Localization of mobile tags essentially follows the same design principles as Section~\ref{sec:subcm_loc}, where mobile tag signal $f_m$ is eliminated to run IFFT (Figures~\ref{fig:SR}(a)-(c)), reconstructing the range frequency in the time domain (i.e., $f_r(t)$).
Subsequently, each subset with duration $T$ of $f_r(t)$ can be zero-padded to reveal the precise location at the corresponding time (Figures~\ref{fig:SR}(d),(e)).
\ours mobile localization can be configured for balance between time granularity and computation overhead, by choosing the location update interval. For instance, zero-padding can be applied on $f_r(t)$ with a $520~T$ interval to provide $60$ localization updates per second (under $T=32~\mu s$).
Our evaluations show $2.6~mm$ median error for a humanoid robot with $17~cm/s$ speed, providing evidence for mobile tag localization \ours.
We note that the minor modification of \ours sustains $O(N\log N)$ computation complexity for mobile localization.

\begin{figure}[h!]
\centering
% \vspace{-0.3cm}
\subfigure[IF for 5 tags]{\includegraphics[width=0.49\columnwidth]{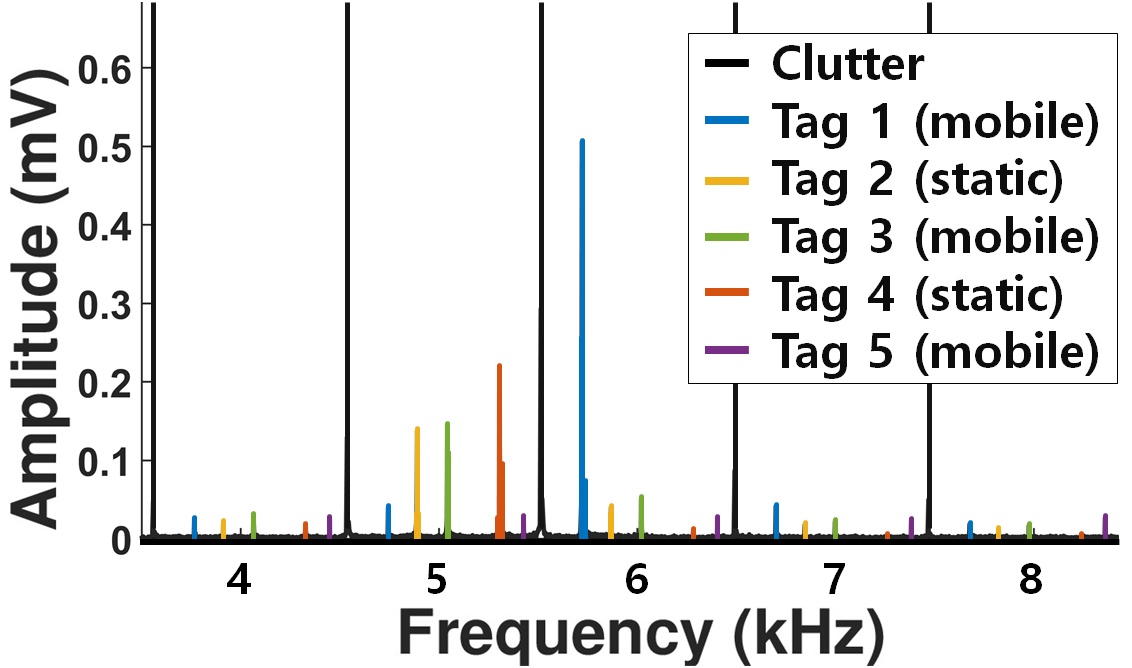}}
\subfigure[Localization]{\includegraphics[width=0.49\columnwidth]{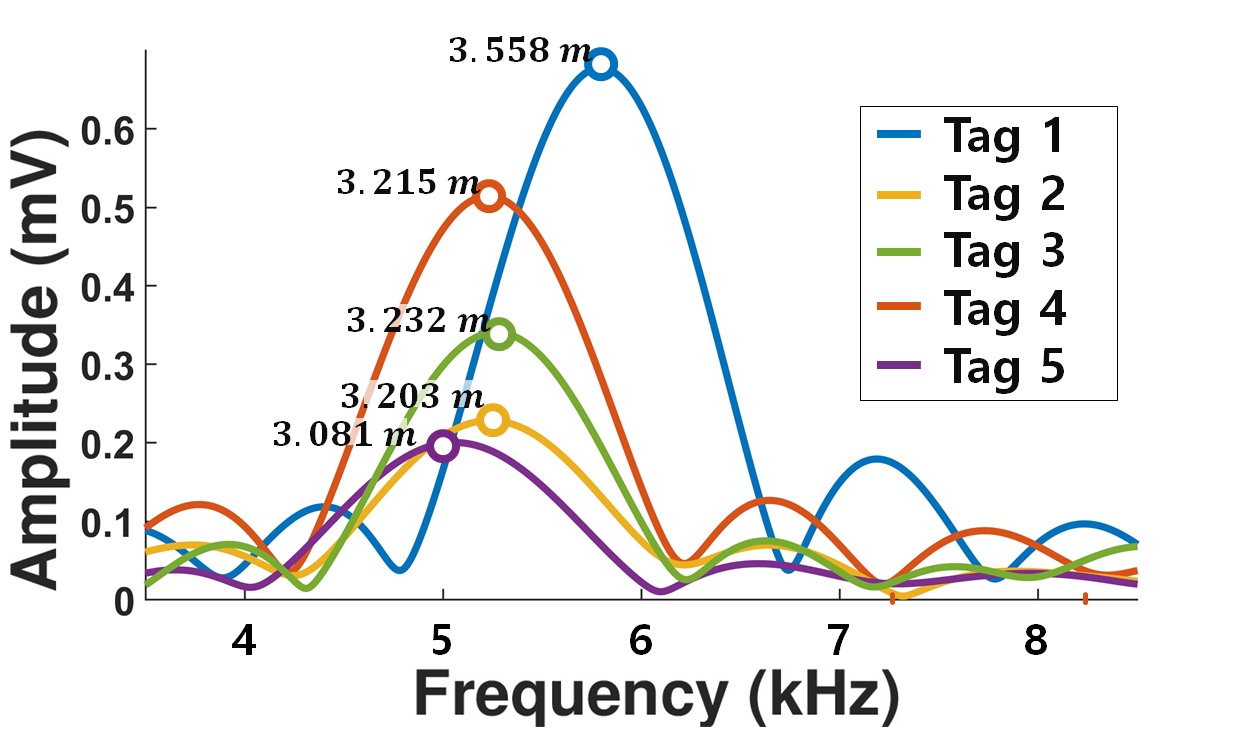}}
\vspace{-4mm}	
\caption{Five tags (a) IF from a single interrogation and (b) its localization results.}\label{fig:multiple_tags}
\vspace{-0.6cm}
\end{figure}

\subsection{Large-scale One-shot Localization}\label{sec:large_scale}
% \vspace{-2mm}

The lightweight localization of \ours can be directly extended to large-scale, for simultaneous localization of mobile and static \ours tags with a single interrogation. For instance, localizing 100 tags takes less than $33.2~ms$ end-to-end ($3.2~ms$ interrogation + $30~ms$ processing time) on a mediocre desktop PC (i7-8700, $32~GB$ RAM).
% up to over thousand tags simultaneously with a single interrogation.
% For instance, localizing 100 tags takes less than $694~ms$ end-to-end, including the time for the interrogation and the computation. The large scalability, in combination with long-range localization and omnidirectional tag, makes \ours especially effective for wide-area coverage.
% \vspace{-0.3mm}
We verify simultaneous localization of 100 tags in Section~\ref{sec:large_eval}, where each tag is identified according to its unique modulating frequencies.
The localization runs iteratively for each tag, to eliminate tag modulation before extracting accurate range frequency.
Figure~\ref{fig:multiple_tags}(a) illustrates simultaneous localization of 5 mobile and static tags, where the tags with 200, 500, 890 Hz modulation are mobile and 350, 770 Hz modulation are static.
Figure~\ref{fig:multiple_tags}(b) depicts successful localization of each tag, where individual tag signals are distinguished according to the modulation frequency.
We note that \ours supports up to 1024 tags under $32.8~ms$ interrogation signal ($T=32~\mu s$ and $N=1025$), which translates to $30.5~Hz$ interval between each tag IDs. The ample ID space tolerates over $500~ppm$ frequency offset in low-end crystal oscillators, demonstrating the scalability of \ours in practice.
The large scalability, in combination with long-range localization and retro-reflective tag, offers a wide-area coverage.

\subsection{Radar Setup}
\vspace{1mm}\noindent{\textbf{Multilateration.}}
\ours is capable of supporting seamless 2D/3D tag localization, where multiple \ours radars concurrently interrogate \ours tags for multilateration.
% Multiple \ours radars can concurrently interrogate \ours tags without collision, 
% \ours achives 2D/3D localization utilizing multiple radars for multilateration, where $8.7~mm$ 2D localization error is achieved at $100~m$ distance in practice.
% \ours tags can be concurrently interrogated in the multilateration process, exempting the necessity for time division between \ours radars.
Essentially, concurrent interrogation is made possible by \ours plannar VAA tag which retro-reflects interrogation signal back to the source radar, efficiently avoiding tag signal interference amongst radars. Hence, \ours radars can be set up for multilateration without the need for access control.
% , and
% i) the HD-FMCW tag signal isolation, which applies generally to chirps from own radar but also to other radars.
% (ii) the tag signal isolation of HD-FMCW where interference from different radars are handled identical to clutter noise, leaving the tag FSK signal intact.
Furthermore, \ours radars can be time-synchronized to support 2D/3D localization for mobile tags, utilizing the Network Time Protocol~\cite{NTP}. The protocol provides sub-millisecond accuracy in local area networks, where a millisecond error translates to $3.6~mm$ localization error for a typical human running speed of $13~kmph$, sustaining subcentimeter accuracy.

\vspace{1mm}\noindent{\textbf{Single Radar Localization.}} Single radar 2D/3D localization can be achieved by utilizing the AoA of the MIMO radar. Compared to multilateration, single radar localization trades off accuracy for lower deployment cost (less number of radars). Localization error from the AoA inaccuracy is amplified over distance. For instance, an AoA error of $5^\circ$ causes $8.7~m$ localization error at $100~m$ ($100~m\times 0.087~rad$). Single radar localization is demonstrated in Section~\ref{sec:2D_eval}.

\begin{figure}[h!]
\centering
\vspace{-0.2cm}
\includegraphics[width=0.8\columnwidth]{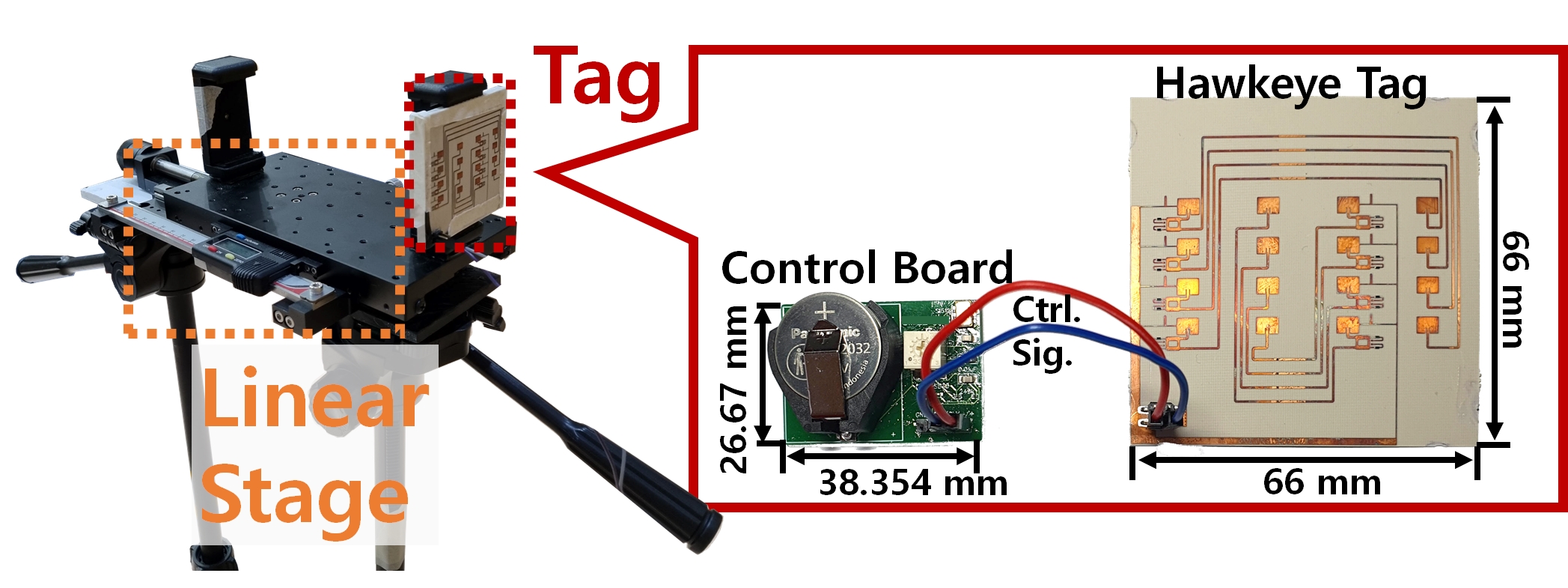}
\vspace{-0.4cm}
\caption{\ours tag evaluation setup. We utilize a separate control board to provide the control signal to tag. The tag is mounted on a linear stage with $0.01~mm$ resolution.}\label{fig:tag_imp}
\vspace{-0.2cm}
\end{figure}

\vspace{-4mm}
\section{Evaluation} \label{sec:eval}
\vspace{-1mm}
This section presents the implementation details and evaluation results of \ours.
\vspace{-1mm}
\subsection{Implementation}
\vspace{-1mm}

% We implement \ours localization on a $24~GHz$ commodity radar. Additionally, two types of control board is implemented for control signal of \ours tag. 

\ours radar is implemented on Eval-TinyRad (Analog Devices)~\cite{tinyrad} commodity 24GHz radar, where the operation of \ours localization is verified. The radars provide the IF data to PC, where it is collected to perform \ours localization. 
% \begin{figure}[h!]
% \centering
% \vspace{-0.1cm}
% \includegraphics[width=1\columnwidth]{figures/tags.jpg}
% % \vspace{-6.5mm}	
% \caption{We utilize a custom control board fabricated using VCXO, which has a small form factor of $26.67~mm \times 38.354~mm$}\label{fig:vcxo}
% \vspace{-0.3cm}
% \end{figure}
% \vspace{0.1in}\noindent \textbf{Control Board.}
To deliver the control signal to \ours tag, a fabricated control board with VCXO (i.e., Voltage Controlled Crystal Oscillator) is used with a small form factor of $26.67~mm \times 38.354~mm$, as depicted in Figure~\ref{fig:tag_imp}. The board uses Skyworks 515NDAM 134200BAG~\cite{vcxo} oscillator for an accurate $f_m$ generation with $20~ppm$ variance. To control the frequency of the VCXO, a variable resistor is combined with a coin cell battery, where a LDO voltage regulator (Toshiba TAR5SB33~\cite{regulator}) is utilized to stabilize the voltage. Arduino Uno is also implemented as a control board for large-scale localization, where it provides a wider range of $f_m$ ($4~\mathrm{MHz}$ bandwidth, $0-4~\mathrm{MHz}$) compared to the VCXO ($25.2~\mathrm{Hz}$ bandwidth, $134.1748-134.2252~\mathrm{kHz}$).

\begin{figure}[h!]
\centering
\vspace{-0.25cm}
\includegraphics[width=0.85\columnwidth]{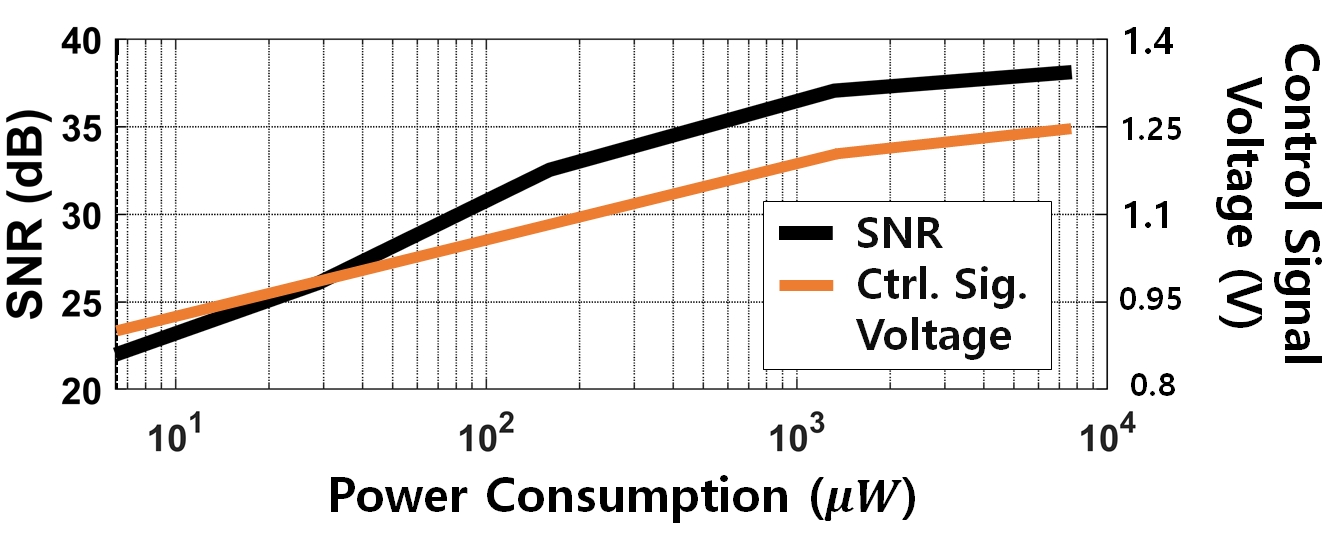}
\vspace{-0.35cm}	
\caption{The power consumption (excluding the control board power) is plotted versus the SNR at $2.4~m$ tag-radar distance with $6.45~dBm$ transmit power. The power consumption ranges from $6.4~\mu W$ to $7.68~mW$.}\label{fig:power}
\vspace{-0.2cm}
\end{figure}

\begin{figure}[h!]
\centering
\vspace{-0.25cm}
\subfigure[]{\includegraphics[width=0.3\columnwidth]{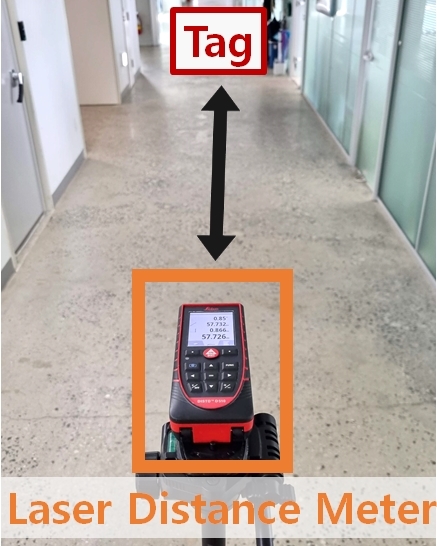}}
\hspace{9mm}
\subfigure[]{\includegraphics[width=0.3\columnwidth]{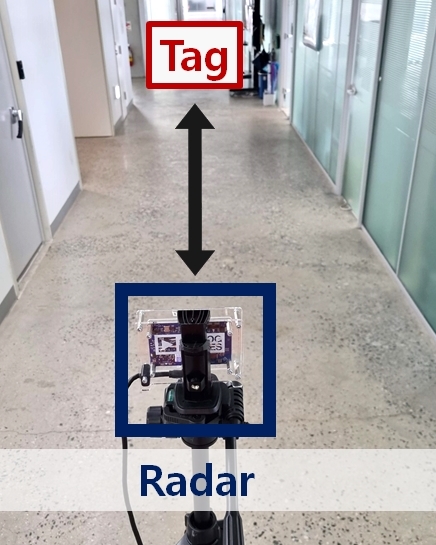}}
\vspace{-4mm}	
\caption{For the ground truth of localization, we utilize a (a) laser distance meter with $1~mm$ resolution mounted on a tripod, then (b) mount the radar on the tripod for evaluation.}\label{fig:gt_imp}
\vspace{-0.65cm}
\end{figure}

% This part is for Evaluation Figure

\vspace{0.1in}\noindent \textbf{Tag Power Consumption.} \ours tag is composed of 16 PIN diodes (Macom MADP-000907-14020), with a separated control board for operation. The tag power consumption is highly variant according to the control signal voltage, as the diodes consume more power at higher control voltage. Figure~\ref{fig:power} provides our evaluation on the power consumption vs. SNR at $2.4~m$ tag-radar distance with $6.45~dBm$ interrogation signal. The power consumption is calculated utilizing the diode IV data~\cite{diodeIV}. The results demonstrate the control signal voltage of $0.9~V$ (i.e., tag operating at $6.4~\mu W$ power) is sufficient for the operation of \ours tag at $2.4~m$ distance with over $20~dB$ SNR.
Meanwhile, the control board power consumption is analyzed by simulating an IC using the Libero SoC SmartPower~\cite{smartpower}. The simulation consists of a ring oscillator and a modulator circuit, where the power consumption results in $2~\mu W$. Thus, \ours tag can run with $8.4~\mu W$ power, which can be easily operated by energy harvesting~\cite{zhang2016hitchhike}, or with a coin cell battery of 1000 $mAh$ for 40.7 years.

\begin{figure}[h!]
    \centering
    
    \subfigure[The tag deployment at \ours 1D localization experiment, performed over $180~m$ at every $20~m$.]{\includegraphics[width=1\columnwidth]{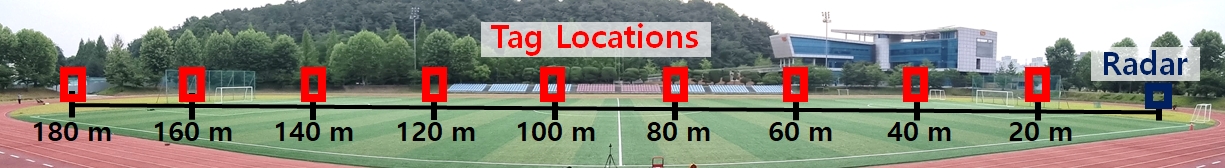}}
    \vspace{-3mm}
    \subfigure[Box plot of \ours 1D localization result.]{\includegraphics[width=0.9\columnwidth]{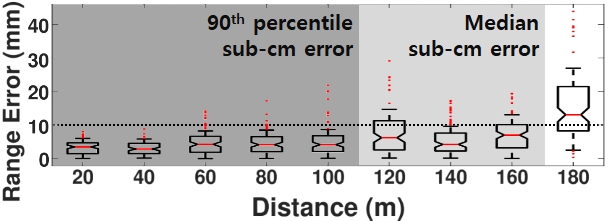}}
    \hspace{5mm}
    \subfigure[Detection rate of \ours. \ours tags achieve stable localization up to $160~m$.]{\includegraphics[width=0.9\columnwidth]{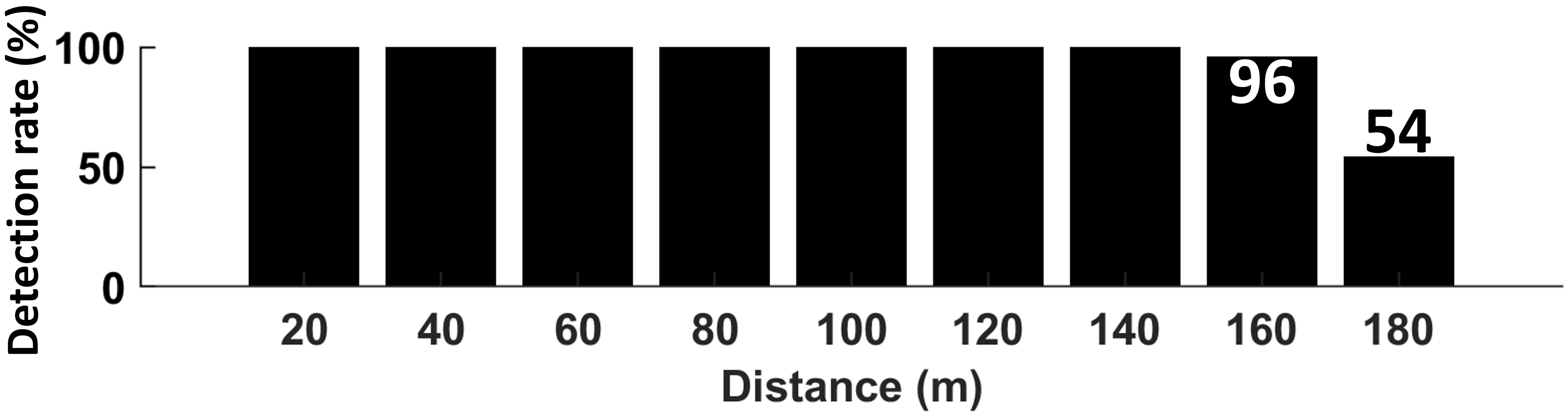}}
    \vspace{-2mm}
    \caption{The 1D localization evaluation setup and performance.}
    % \textbf{\caption{The SNR boost and tag signal demodulation with and without HD-FMCW.}}
    \label{fig:1d}
    \vspace{-5mm}
\end{figure}

\vspace{-1mm}
\subsection{Evaluation settings}
\vspace{-0.5mm}
By default, we conduct evaluations using Analog Devices Eval-Tinyrad as \ours radar, with the specific radar parameters set as follows: bandwidth 250 MHz (24 GHz to 24.25 GHz), transmit power $8~dBm$, IF sampling frequency 1 MHz, and 8192 samples per chirp with 2048 chirps for interrogation signal. We use a single transmit antenna and a single receive antenna for omni-directional \ours operation, unless otherwise mentioned. We note that the PCB fabrication and soldering error may result in separate distance offset per tag,
% the radar oscillator is affected by temperature change~\cite{downconverter},
which we calibrate by measuring the distance offset of a tag with known distance.
In all evaluations, \ours tag is mounted on a acrylic plate to avoid unnecessary electric coupling. The Arduino Uno is utilized for our control board, which supplies $1.3~V$ control voltage to \ours tag.

\vspace{1mm}\noindent \textbf{Ground Truth.}
In order to obtain precise ground truth of localization, laser distance meter with $1~mm$ resolution up to 200 meters range (Leica DISTO D510~\cite{laser}) is utilized at radar side, as depicted in Figure~\ref{fig:gt_imp}. The tag-radar distance is first measured on the laser distance meter mounted on a tripod, where subsequently \ours radar is mounted on the same tripod for evaluation. At tag side, a linear stage of $0.01~mm$ resolution up to $150~mm$ range (Soar STMX1020-D~\cite{stage}) is utilized as shown in Figure~\ref{fig:tag_imp}. \ours tag is mounted on the linear stage, where the linear stage relocates the tag for evaluation. The ground truth error caused by the laser distance meter does not exceed $1~mm$. For mobile tag evaluation, we utilize OptiTrack PrimeX~\cite{optitrack} with $0.2~mm$ 3D accuracy for ground truth.

% Evaluation precise measurement is by ~\ref{fig:gt}. This calibration

\vspace{-1mm}
\subsection{1D Localization}\label{sec:lDlocalize}
% \vspace{-1.5mm}

To verify \ours's subcentimeter accuracy at hectometer range, a 1D localization experiment is conducted at a soccer field, where the measurement is conducted up to 180 meters in a straight line. \ours tag is located at every 20 meters as shown in Figure~\ref{fig:1d}(a), where a total of 100 experimental trials are performed at 20 different locations within each $20~m$ position. Figure~\ref{fig:1d}(b) demonstrates \ours 1D localization performance, where the edge of the box indicates the $75^{th}$ and $25^{th}$ percentile error, while the whiskers indicate the $90^{th}$ and $10^{th}$ percentile error. Data outside the $90^{th}$ and $10^{th}$ percentile error is considered as outliers, which are marked as red dots outside the whiskers. \ours achieves subcentimeter accuracy with $90^{th}$ percentile error up to $100~m$, where the error is $8.9~mm$. Furthermore, the subcentimeter accuracy is sustained up to $160~m$ with $50^{th}$ percentile (i.e., median) error, where the median error is $6.7~mm$ at $160~m$, demonstrating successful hectometer range subcentimeter localization. For control signal, Arduino Uno produces $150~\mathrm{kHz}$ $f_m$. Figure~\ref{fig:1d}(c) shows the detection rate of \ours, where we achieve $100~\%$ detection rate up to $140~m$, which is decreased to $96~\%$ and $54~\%$ at $160~m$ and $180~m$. The subcentimeter accuracy at hectometer range proves the robustness of \ours tag, in combination with the efficiency of \ours localization.

\begin{figure}[h!]
\centering
\vspace{-0.1cm}
\includegraphics[width=\columnwidth]{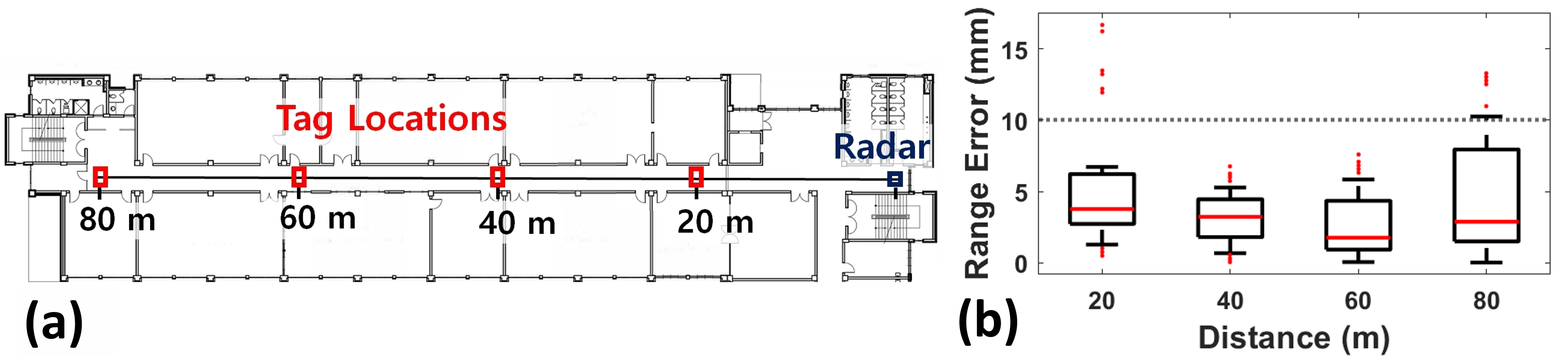}
\vspace{-6mm}	
\caption{(a) Indoor hallway experiment setup of \ours, where the tag is placed at every $20~m$ from the radar.
(b) Box plot of hallway experiment results.
}\label{fig:1d_indoor}
\vspace{-0.2cm}
\end{figure}

% \begin{figure}[h!]
% \centering
% \vspace{-0.1cm}
% \includegraphics[width=0.9\columnwidth]{figures/box_in.jpg}
% \vspace{-1mm}	
% \caption{Box plot of hallway experiment results.\hankyeol{place to side by side}}\label{fig:indoor_box}
% \vspace{-0.4cm}
% \end{figure}

\vspace{1mm}\noindent \textbf{Indoor 1D Localization.}
We further evaluate 1D localization at multipath rich hallway to verify \ours operation in indoors. As depicted in Figure~\ref{fig:1d_indoor}(a), the hallway experiment is conducted up to $80~m$ in a straight line. The tag is located at every $20~m$, where 50 experimental trials are conducted at 10 different locations within each $20~m$ positions. Figure~\ref{fig:1d_indoor}(b) plots the box plot of the indoor 1D localization. The median error stays below $4~mm$ throughout 80 meters, validating \ours's subcentimeter localization even in indoors. The detection rate stayed at $100~\%$ up to the $80~m$ range. This results totally prove the multipath suppression of \ours tag.

\begin{figure}[h!]
\centering
\vspace{-0.15cm}
{\includegraphics[width=0.9\columnwidth]{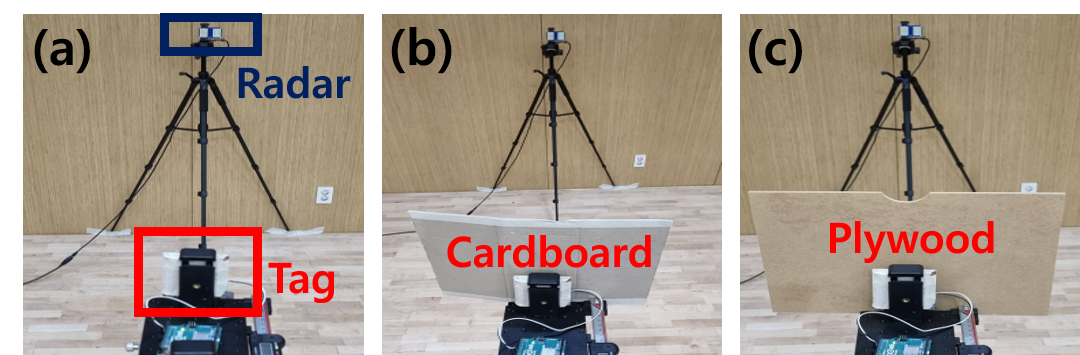}}
\vspace{-1mm}	
\caption{Experiment setup for testing the effect of blockage in the indoor environment (a) without blockage (LOS) (b) with cardboard (c) with plywood. }\label{fig:blockage}
\vspace{-0.45cm}
\end{figure}

\begin{center}
\vspace{-0.15cm}
\begin{tabular}{|c||c|c|c|}  %<-- change here
\hline 
 & LOS & Cardboard & Plywood \\ 
\hline 
\hline 
Median (mm) & 2.2 & 3.4 & 5.7 \\ 
\hline 
$90^{th}$ percentile (mm) & 5.9 & 5.7 & 11.9 \\ 
\hline 
\end{tabular}
\vspace{-2mm}
\captionof{table}{1D localization accuracy under blockages.} 
\vspace{-3mm}
\label{table:blockage}
\end{center}

\begin{figure}[h!]
\centering
{\includegraphics[width=0.9\columnwidth]{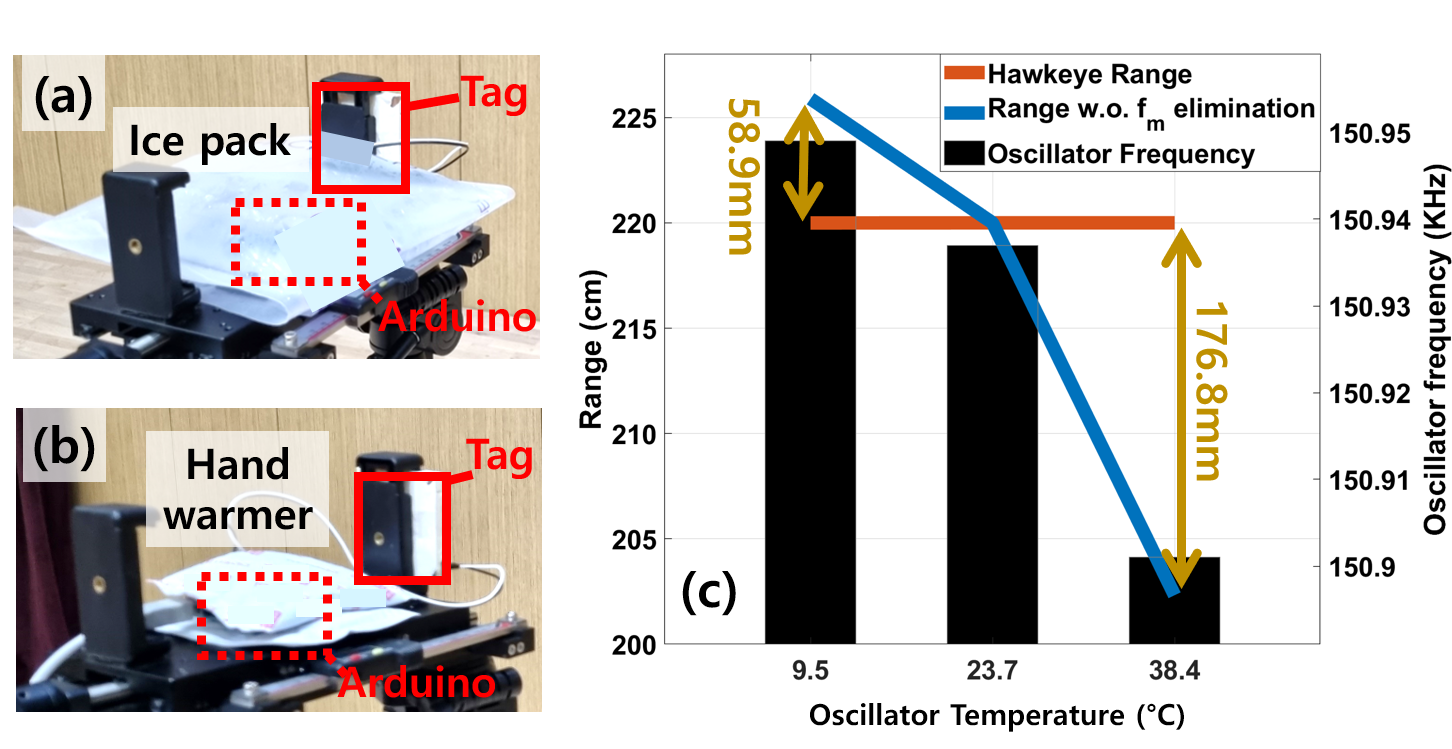}}
\vspace{-3mm}
\caption{Experiment setup for testing the effect of temperature at indoor environment with (a) ice pack (b) hand warmer. (c) \ours effectively eliminates the effect of the instability of oscillator.}\label{fig:temperature}
\vspace{-5mm}
\end{figure}

% \begin{figure}[h!]
% \centering
% \vspace{-0.1cm}
% {\includegraphics[width=\columnwidth]{figures/range_vs_freq.png}}
% % \vspace{-6.5mm}	
% \caption{asdasd }\label{fig:range_freq}
% \vspace{-0.3cm}
% \end{figure}

\vspace{0.1in}\noindent \textbf{Robustness Against Blockage and Temperature Change.}
Localization for IoT applications may face various environment changes during practical use. In order to substantiate \ours operation for pervasive deployment, we analyze the 1D localization performance impact of two common error sources -- blockages and temperature variance. 
% we note that temperature change can impact the oscillator performance~\cite{arduino_osc}, where our experiment showed median frequency variance of xx at yy temperature change.
The experiments are conducted in an indoor concert hall, where the tag-radar distance is set to $2.2~m$.
For each experiment, total 40 experimental trials are conducted at 4 different locations, with Arduino Uno as the control board. 
% The control board at room temp with LOS is provided as reference.
Figure~\ref{fig:blockage} shows the setup for the blockage experiments, where cardboard and plywood of $3.25~mm$ and $5~mm$ thickness is utilized as blockage materials.
The experiment results are summarized at Table~\ref{table:blockage}, where the cardboard and plywood blockage increased the median localization error by $1.2~mm$ and $3.5~mm$ each. The results show that \ours sustains subcentimeter accuracy even under NLOS, demonstrating our robustness to the blockages.
Localization performance under varying temperature (i.e., varying $f_m$ due to the instability of oscillator) is also evaluated, to show \ours's ability to eliminate the effect of $f_m$ for precise localization. As shown in Figure~\ref{fig:temperature}, the Arduino oscillator temperature is set to $9.54^\circ C$, $23.7^\circ C$ and $38.43^\circ C$, for evaluation. At each temperature, the $f_m$ varied from $150.901~\mathrm{kHz}$ to $150.949$ $\mathrm{kHz}$, showing high frequency instability of the control board. To control the temperature of the oscillator, the control board was either surrounded by ice pack or hand warmers, as shown in Figures~\ref{fig:temperature}(a) and (b).
Figure~\ref{fig:temperature}(c) compares the localization result with and without \ours $f_m$ elimination. Without the $f_m$ elimination, the error induced by the temperature at $9.54^\circ C$ and $38.43^\circ C$ is $58.9~mm$ and $176.8~mm$ each. Contrarily, with \ours's $f_m$ elimination applied, the error induced by the temperature change at $9.54^\circ C$ and $38.43^\circ C$ stays under $4~mm$.
Altogether, the results prove the performance of \ours at diverse environments.

% \begin{figure}[h!]
%     \centering
%     \vspace{-2mm}
%     \includegraphics[width=0.8\columnwidth]{figures/2d2.jpg}
%     \vspace{-1mm}
%     \caption{The experimental setup of 2D localization experiment. The tag and two radars form an equilateral triangle with a side of $100~m$.}
%     % \textbf{\caption{The SNR boost and tag signal demodulation with and without HD-FMCW.}}
%     \label{fig:2d}
%     \vspace{-3mm}
% \end{figure}

% \begin{figure}[h!]
%     \centering
%     % \vspace{-4mm}
%     \includegraphics[width=0.8\columnwidth]{figures/2d_3.jpg}
%     % \vspace{-7.5mm}
%     \caption{The CDF of \ours 2D localization error, where subcentimeter median error is achieved.}
%     % \textbf{\caption{The SNR boost and tag signal demodulation with and without HD-FMCW.}}
%     \label{fig:2dcdf}
%     \vspace{-3mm}
% \end{figure}

\begin{figure}[h!]
    \centering
    \vspace{-2mm}
    \includegraphics[width=\columnwidth]{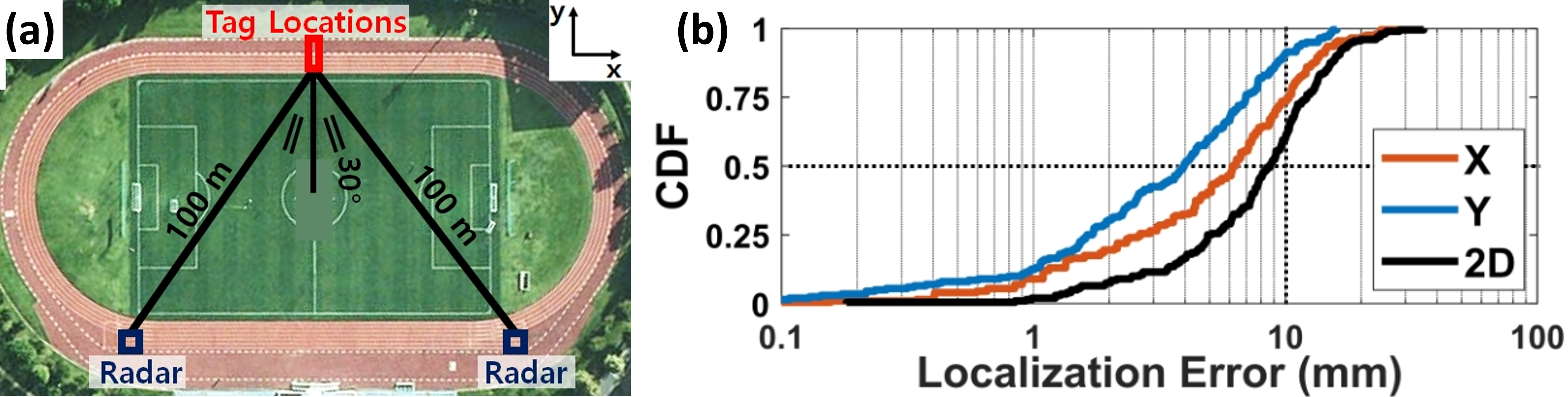}
    \vspace{-4mm}
    \caption{(a) The experimental setup of 2D localization experiment. The tag and two radars form an equilateral triangle with a side of $100~m$. (b) The CDF of \ours 2D localization error.}
    % \textbf{\caption{The SNR boost and tag signal demodulation with and without HD-FMCW.}}
    \label{fig:2d}
    \vspace{-3mm}
\end{figure}

\vspace{-1mm}
\subsection{2D Localization}\label{sec:2D_eval}
% \vspace{-1.5mm}
We evaluate hectometer range 2D localization at soccer field with track, utilizing two radars and a single tag for multilateration. As demonstrated at Figure~\ref{fig:2d}(a), the two radars are located at $100~m$ distance from the tag, where the radar interrogates the tag with $30^\circ$ incidence angle. The distance between the radars is $100~m$, and the tag modulation $f_m$ is 150 kHz. The experiment consists of total 400 experimental trials, where the tag localization is conducted at 20 positions on the linear stage moving towards positive $y$ direction. Figure~\ref{fig:2d}(b) plots the CDF of 2D localization error, where \ours achieves subcentimeter median error of $8.7~mm$ and $90^{th}$ percentile accuracy of $15.9~mm$ in 2D. Median error along $x$ and $y$ dimensions are $6.2~mm$ and $4~mm$, while the $90^{th}$ percentile error is $13.6~mm$ and $9.8~mm$ each. The 2D localization, which inherently requires retro-reflectivity, is made possible by \ours tag's high retro-reflectivity. The subcentimeter 2D localization at hectometer range demonstrates the capability of \ours at practical applications.

\begin{figure}[h!]
\centering
\vspace{-4mm}
\includegraphics[width=\columnwidth]{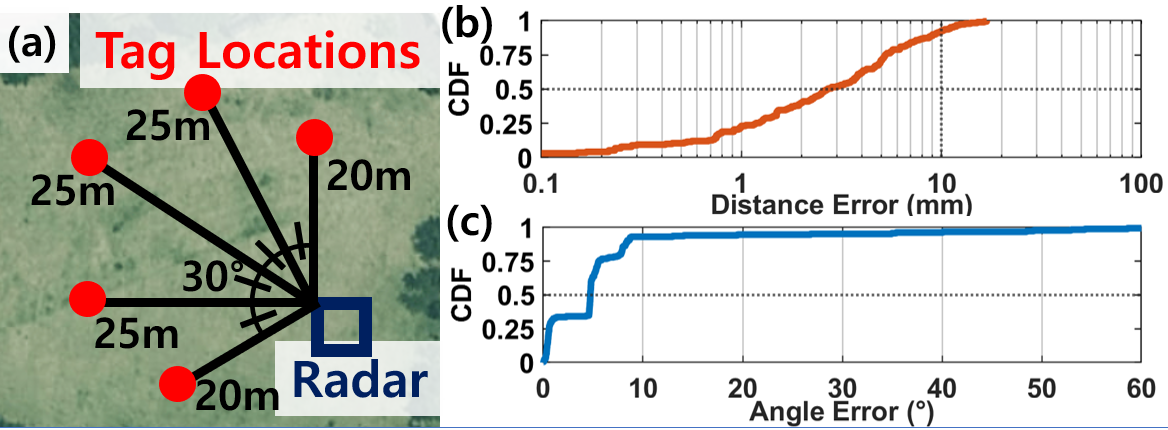}
% \vspace{-7.5mm}
\caption{The CDF of \ours 2D localization error with single radar.}
% \textbf{\caption{The SNR boost and tag signal demodulation with and without HD-FMCW.}}
\label{fig:2daoa}
\vspace{-2mm}
\end{figure}

\vspace{0.1in}\noindent{\textbf{Single Radar 2D Localization.}}
\ours 2D localization on a single radar utilizing AoA is evaluated. The evaluation is conducted at an open field with localization distance ranging from $20$ to $25~m$ and AoA ranging from $-60^\circ$ to $60^\circ$, as depicted in Figure~\ref{fig:2daoa}(a). For each location, a total of 30 experimental trials are performed at 6 different locations. Figure~\ref{fig:2daoa}(b) and (c) show the CDF of distance and angle error, where the median errors are $2.7~mm$ and $4.7^\circ$, and the $90^{th}$ percentile errors are $9.04~mm$ and $8.4^\circ$.
% \ours helps localization in the 2D plane with single radar, by enhancing the distance accuracy. 
Collectively, the median 2D error results in $2.08~m$, whose accuracy can be improved with a larger antenna array.

\begin{figure}[h!]
    \vspace{-4mm}
    \centering
    \subfigure[3D setup.]{\includegraphics[width=0.3\columnwidth]{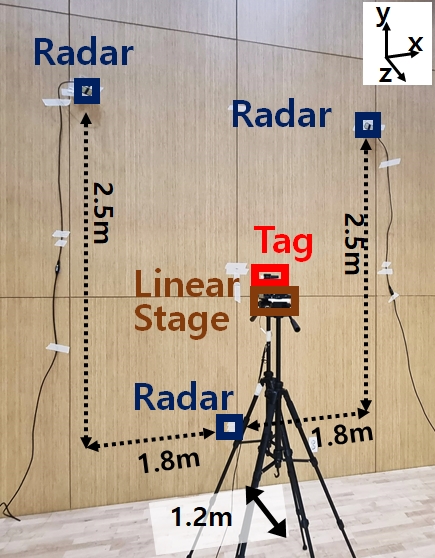}}
    \hspace{3mm}
    \subfigure[Localization result.]{\includegraphics[width=0.6\columnwidth]{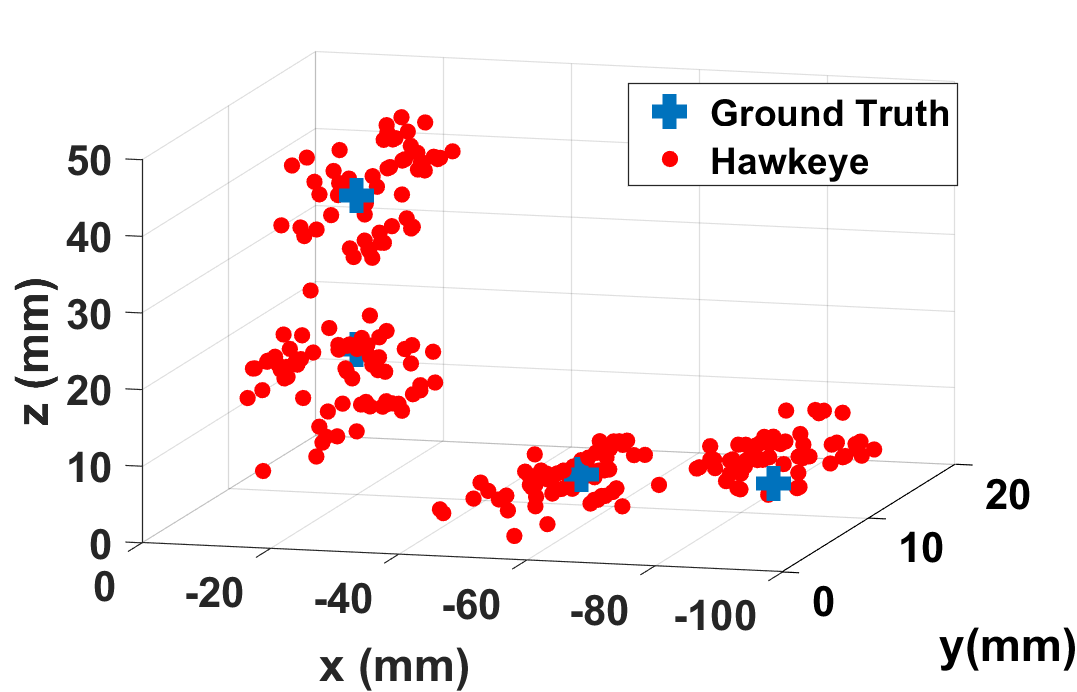}}
    \vspace{-3mm}
    \caption{(a) Experiment setup of 3D localization experiment, conducted within a indoor concert hall. (b) An exemplary localization result at four tag locations, where the estimated points within the median error are plotted.}
    % \textbf{\caption{The SNR boost and tag signal demodulation with and without HD-FMCW.}}
    \label{fig:3d}
    \vspace{-4mm}
\end{figure}

\begin{figure}[h!]
    \centering
    \includegraphics[width=0.95\columnwidth]{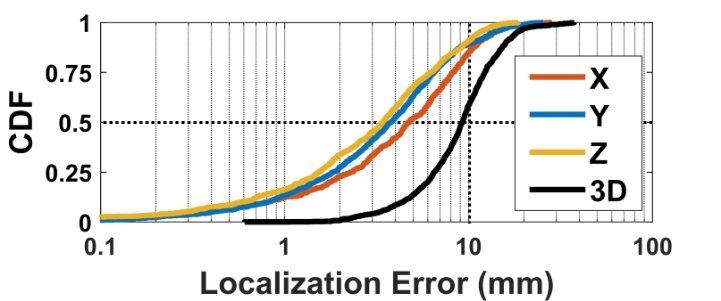}
    \vspace{-3mm}
    \caption{The CDF of 3D localization error.
    % Subcentimeter median error is achieved even at indoor 3D localization.
    }
    % \textbf{\caption{The SNR boost and tag signal demodulation with and without HD-FMCW.}}
    \vspace{-2mm}
    \label{fig:3dcdf}
\end{figure}

% \begin{figure}[h!]
%     % \vspace{-5mm}
%     \centering
%     % \vspace{2mm}
%     \includegraphics[width=0.8\columnwidth]{figures/pointcloud_ratiochange.png}
%     \vspace{-1mm}
%     \caption{An exemplary localization result at four tag locations, where the estimated points within the median error are plotted.}
%     % \textbf{\caption{The SNR boost and tag signal demodulation with and without HD-FMCW.}}
%     \label{fig:3d_pointcloud}
%     \vspace{-5mm}
% \end{figure}

\vspace{2mm}
\subsection{3D Localization}
\vspace{-1mm}
In order to verify the 3D localization performance of \ours, a localization experiment is conducted in an indoor concert hall with three radars fixed to a wall.
% We carefully position the radars to maximize the localization accuracy by forming right angle at each radar-tag-radar angle.
The radar positions are described in Figure~\ref{fig:3d}(a), where the three radars are located at (1.8, 1.25, -1.2), (1.8, -1.25, -1.2) and (0, -1.25, -1.2) coordinates, assuming $(0,0,0)$ is the tag center. Tag control signal of $1.15~V$ is fed with a signal generator~\cite{siggen} for 3D localization experiment. Total 1840 experimental trials at 23 positions in $xz$-plane are conducted utilizing the linear stage. Figure~\ref{fig:3dcdf} demonstrates the CDF of 3D localization error, presenting subcentimeter median error of $8.9~mm$ and $90^{th}$ percentile error of $13.8~mm$. For each $x,y,z$ dimensions, the median error is $5.9~mm$, $2.7~mm$, $3.4~mm$ and $90^{th}$ percentile error is $11.2~mm$, $8.5~mm$, $8.8~mm$ each. Figure~\ref{fig:3d}(b) depicts an exemplary localization result at four tag locations, where estimated points within the median error are plotted in 3D. The successful subcentimeter 3D localization in indoor space proves the retro-reflective performance of \ours tag in both azimuth and elevation plane, while establishing a solid foundation on the practicality of \ours localization.

\begin{figure}[h!]
    % \vspace{-5mm}
    \centering
    % \vspace{2mm}
    \includegraphics[width=1\columnwidth]{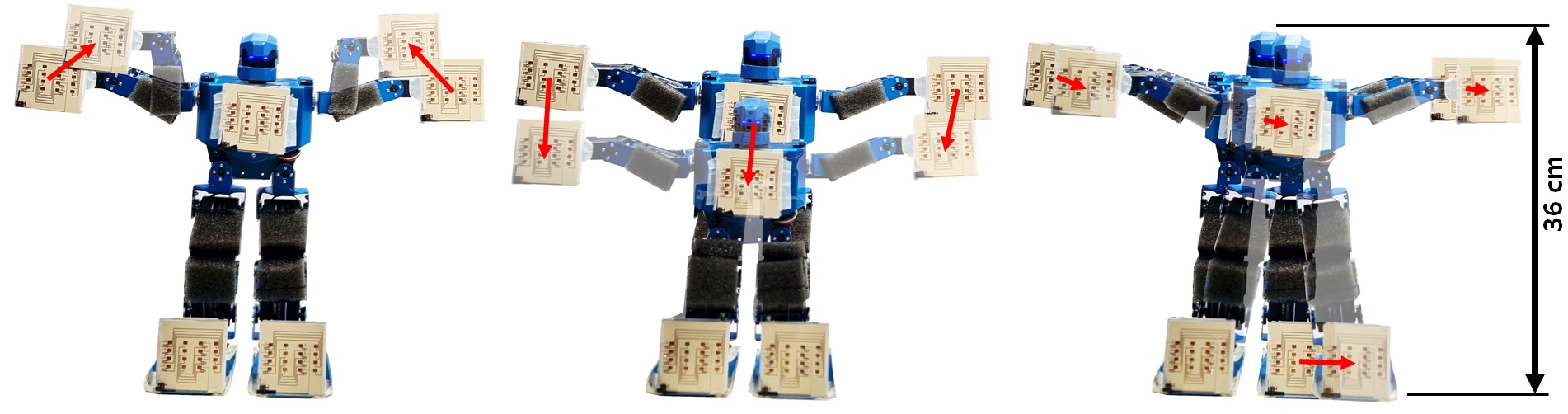}
    \includegraphics[width=1\columnwidth]{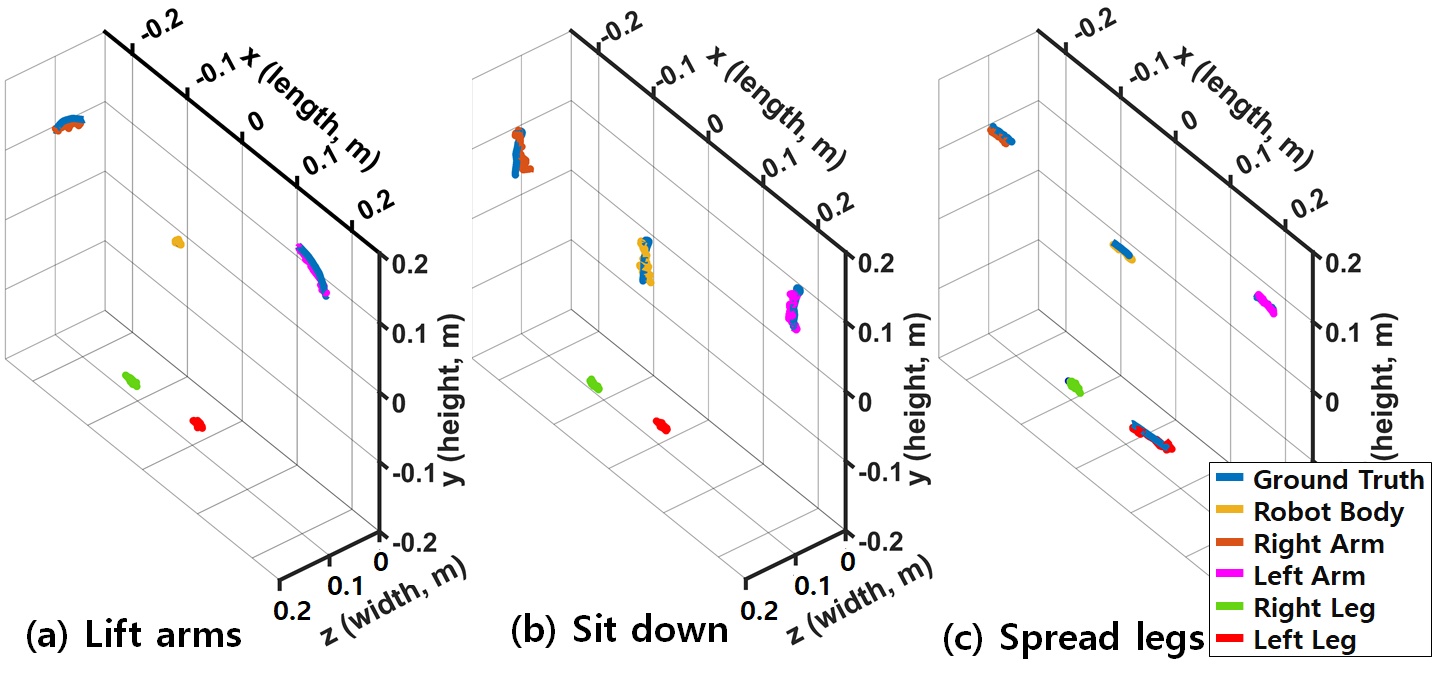}
    \vspace{-5mm}
    \caption{Robot movements and the corresponding localization results.}
    % \textbf{\caption{The SNR boost and tag signal demodulation with and without HD-FMCW.}}
    \label{fig:robotmoves}
    \vspace{-5mm}
\end{figure}

\begin{figure}[h!]
    \centering
    \includegraphics[width=0.95\columnwidth]{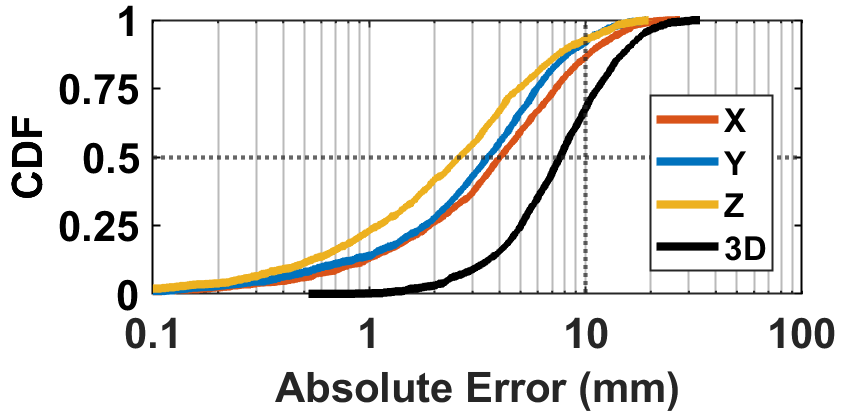}
    \vspace{-3mm}
    \caption{The CDF of 3D localization error from moving robot.
    }
    % \textbf{\caption{The SNR boost and tag signal demodulation with and without HD-FMCW.}}
    \vspace{-5mm}
    \label{fig:robotcdf}
\end{figure}

\vspace{0.1in}\noindent{\textbf{Simultaneous Localization of Mobile Tags.}}
We evaluate \ours's ability to simultaneously localize mobile tags by attaching five \ours tags to the body center, both legs and both arms of a humanoid robot~\cite{h5s}. Each tag concurrently modulates with unique $f_m$ between 150~kHz and 151~kHz. 
The experiment is conducted in an indoor concert hall with the same settings as Figure~\ref{fig:3d}(a).
The robot has dimension of $48\times36 cm$, operating with 16 servo motors. 
The radar is set to have 1024 samples per chirp with 2048 chirps for interrogation signal.
As depicted in Figure~\ref{fig:robotmoves}, three different actions of lift arms, sit down and spread legs are captured with the maximum moving speed of 17cm/s.
A total of 90 experimental trials are conducted per action, with three different robot positions.
As depicted, the three robot actions are captured with 23.5 localization FPS, where a ground truth measured by OptiTrack is provided together.
The CDF is presented in Figure~\ref{fig:robotcdf}, where the median error of 3D localization is $7.66~mm$ and $90^{th}$ percentile error is $15.96~mm$. For each $x,y,z$ dimension, the median error is $4.05~mm$, $3.53~mm$, $2.59~mm$ and $90^{th}$ percentile error is $11.13~mm$, $9.07~mm$, $8.31~mm$ each.
This verifies \ours's one-shot localization of mobile tags in practice.
% \kangmin{Robot specs, dimensions, actions, median error. Moving Speed max 17cm/s, Localization FPS.}

% \begin{figure}[t!]
%     \centering
%     % \vspace{-4mm}
%     \includegraphics[width=0.65\columnwidth]{figures/100_2.jpg}
%     % \vspace{-7.5mm}
%     \caption{The localization result of 100 tags deployed 10 by 10 on a acrylic board projected on the experiment photo.}
%     % \textbf{\caption{The SNR boost and tag signal demodulation with and without HD-FMCW.}}
%     \label{fig:10by10}
%     \vspace{-3mm}
% \end{figure}

% \begin{figure}[t!]
%     \centering
%     % \vspace{-4mm}
%     \includegraphics[width=0.85\columnwidth]{figures/massive_pointcloud_vert_final.png}
%     \vspace{-3mm}
%     \caption{The accurate localization result of \ours at scale, where all estimation results are shown as red dots around the ground truth.}
%     % \textbf{\caption{The SNR boost and tag signal demodulation with and without HD-FMCW.}}
%     \label{fig:100cloud}
%     \vspace{-3mm}
% \end{figure}

% All large scale figures

\begin{figure*}[h!]
    \centering
    % \vspace{-4mm}
    \includegraphics[width=2.0\columnwidth]{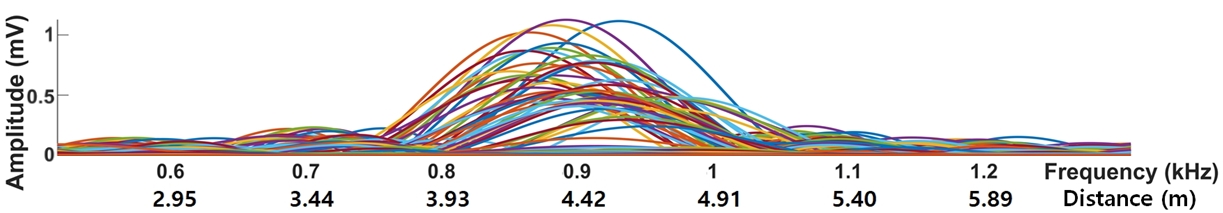}
    \vspace{-2.5mm}
    \caption{Sinc envelope of 100 tags during the large-scale simultaneous localization experiment.}
    % \textbf{\caption{The SNR boost and tag signal demodulation with and without HD-FMCW.}}
    \label{fig:100sincs}
    \vspace{-3.5mm}
\end{figure*}

\begin{figure}[h]
    \centering
    % \vspace{-4mm}
    \includegraphics[width=0.9\columnwidth]{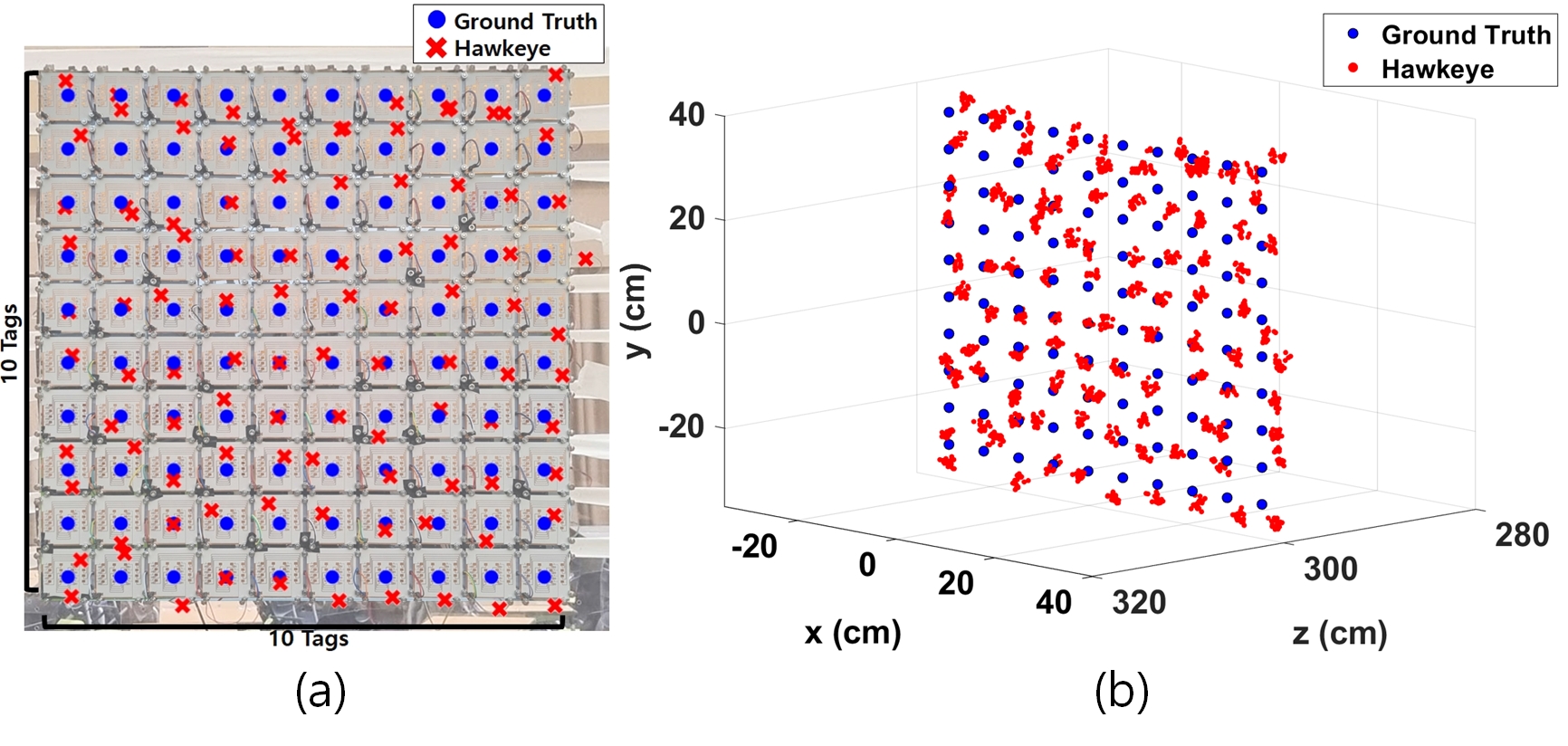}
    \vspace{-1mm}
    \caption{(a) The localization result of 100 tags deployed 10 by 10 on a acrylic board projected on the experiment photo, and (b) the accurate localization result of \ours at scale.}
    % \textbf{\caption{The SNR boost and tag signal demodulation with and without HD-FMCW.}}
    \label{fig:10by10}
    \vspace{-3mm}
\end{figure}

\begin{figure}[h]
    \centering
    % \vspace{-4mm}
    \includegraphics[width=0.8\columnwidth]{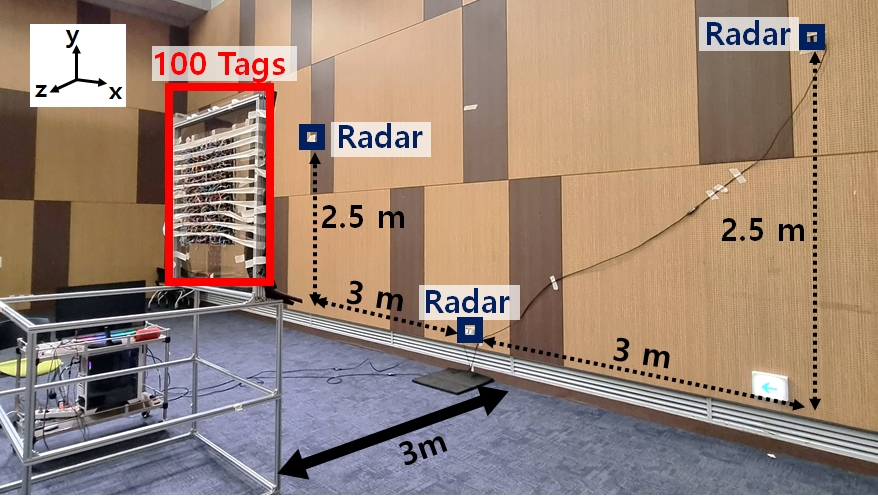}
    \vspace{-1.5mm}
    \caption{Experiment setup of large-scale simultaneous localization. The experiment is conducted indoors, with $5~mm$ spacing between the 100 tags.}
    % \textbf{\caption{The SNR boost and tag signal demodulation with and without HD-FMCW.}}
    \label{fig:100setup}
    \vspace{-2mm}
\end{figure}

\begin{figure}[h!]
    \centering
    % \vspace{-4mm}
    \includegraphics[width=0.95\columnwidth]{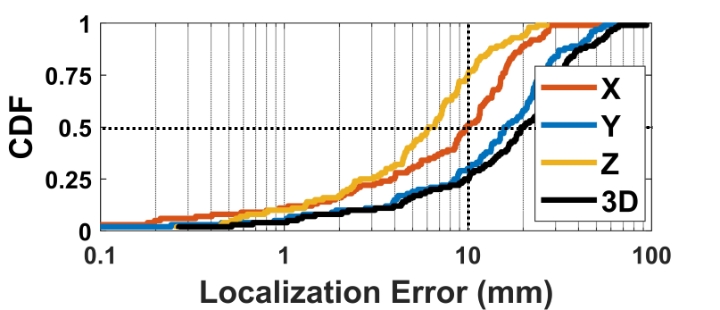}
    \vspace{-3mm}
    \caption{The CDF of large-scale localization error at \ours.}
    % \textbf{\caption{The SNR boost and tag signal demodulation with and without HD-FMCW.}}
    \label{fig:100dcf}
    \vspace{-3mm}
\end{figure}

% \vspace{-3mm}
\subsection{Large-scale Localization}\label{sec:large_eval}
% \vspace{-1mm}
To verify the large-scale support of \ours, a simultaneous, 3D localization experiment consisting of 100 tags is conducted in an indoor environment.
Figure~\ref{fig:10by10}(a) depicts the arrangement of 100 tags, where they are deployed as 10 by 10 on a acrylic board.
The tags are densely deployed with intervals of $5~mm$ to demonstrate operation in harsh environments (e.g., items stacked up in the warehouse) where substantial coupling between tags exists. The 100 tags concurrently modulates with unique $f_m$ in between 100 kHz and 250 kHz. For 3D localization, three radars are attached to a wall as depicted in Figure~\ref{fig:100setup}. Each radars are located at $(3, 1.3, -3)$, $(-3, 1.3, -3)$ and $(0, -1.2, -3)$ coordinates, assuming $(0,0,0)$ is the 100 tags center. Figure~\ref{fig:100sincs} demonstrates the 100 tags localization processing result, where the successful localization of each tags resulting in 100 different envelope sincs are visible.
% Figure~\ref{fig:100cloud}
Figure~\ref{fig:10by10}(b) depicts the the localization result of \ours at scale, where all estimation results are shown as red cross around the ground truth.
The CDF of large-scale 3D localization error is shown at Figure~\ref{fig:100dcf}, showing median error of $19.9~mm$ and $90^{th}$ percentile error of $48.4~mm$. For each $x$, $y$, $z$ dimension, the median error is $9.8~mm$, $16.1~mm$, $6.4~mm$, and $90^{th}$ percentile error is $20.5~mm$, $41.1~mm$, $15.5~mm$ each. The concurrent localization of 100 tags verifies \ours scalability at realistic settings.

\vspace{-1mm}
\section{Related Work}
% \vspace{-1mm}

% \noindent\textbf{Backscatter System.} 
% The literature introduced backscatter to establish an extremely energy-efficient wireless communication links, where \cite{talla2021advances, wang2019rfid} serve good overviews of recent successes. Owing to its lightweight operation, backscatter have been widely installed on low-power tags for diverse sensing applications; localization \cite{chuo2017rf, kempke2016surepoint, luo20193d, ma20163d, ma2017minding_same, nandakumar20183d, pannuto2018slocalization, soltanaghaei2021tagfi, wei2016gyro, xie2020exploring, xu2019faho}, action recognition \cite{gao2018livetag, li2016deep, ryoo2018barnet} and shape sensing \cite{jin2018wish, vasisht2018body}. Backscattered signals can even carry material info \cite{ha2020food, xie2019tagtag}.

\noindent\textbf{mmWave Systems and Sensing.}
mmWave systems have been proposed to exploit the large bandwidth \cite{lacruz2020mm, li2021spider, zhao2020m, haider2018listeer}. mmWave radars also utilize extensive bandwidth to achieve higher sensing accuracy \cite{kong2022m3track, chen2022soil, shuai2021millieye, li2021vocalprint}.

\noindent\textbf{Backscatter.}
Backscatter offers extremely low-power signaling for power constrained scenarios \cite{chi2020leveraging, huang2022rascatter, zhu2022enabling, majid2019multi}. Backscatter system also facilitates low-power sensing, including RFID-based approaches \cite{chang2021rf, wang2017tagscan, nandakumar20183d, li2022back}.

\noindent\textbf{Backscatter/RFID Localization.}
As one of the representative implementation of backscatter systems, RFID-based localization techniques have long been discussed in the community \cite{jiang2018orientation, karmakar2013chipless, shu2015toc, wang2013rf, yang2021rf}. Initial studies measure the amplitude, phase and the angle of arrival of the received signal \cite{arnitz2009multifrequency, azzouzi2011new, li2009multifrequency, zhou2009rfid, kronberger2011uhf, zhou2011two}, which suffered from multi-path effects. 
Recent studies still suffer from low accuracy in practical scenarios due to limited bandwidth of RFID, including \cite{han2015twins, wang2013dude, parr2013inverse, yang2014tagoram, zhang2021siloc} which require knowledge on motion or reference tags to mitigate multi-path effects. 
% Reference tags \cite{bouet2008range, han2015twins, wang2013dude} or prior knowledge on the object’s motion \cite{miesen2011holographic, parr2013inverse, yang2014tagoram, zhang2021siloc} have shown to mitigate the multi-path effect, but have limited practicality. Recent studies still suffer from low accuracy due to limited bandwidth of RFID. 
While \cite{ma2017minding} emulates large bandwidth to achieve sub-centimeter accuracy, range is limited to room-scale scenarios to comply with FCC regulations. \cite{soltanaghaei2021millimetro} utilize large bandwidth of mmWave FMCW radar to achieve $100~m$ range, but is still essentially limited to FMCW range resolution.

\vspace{-1mm}
\section{Conclusion}
% \vspace{-1mm}

This paper presents \ours, a mmWave backscatter localization that can achieve subcentimeter median accuracy at hectometer range, while using an affordable commodity radar ($\sim$200 USD~\cite{distance2go}) for the reader. \ours simultaneously localizes 100 tags in only $33.2~ms$, and is capable of supporting up to 1024 tags in theory. Our design consists of (i) a new planar Van Atta Array tag that retro-reflects in both azimuth and elevation, combined with a low-loss FSK modulator, and (ii) a novel localization algorithm that achieves $\times 60$ the localization performance of FMCW radar, while being immune to the tag oscillator frequency offset. Collectively, \ours take a solid step towards bringing pervasive tag deployment and localization to practice. 

% This paper presents \ours, the first large-scale mmWave backscatter system to exploit the wide mmWave bandwidth. \ours robustly reaches high scalability under practical indoor settings with rich ambient reflections and obstructions. Extensive testbed evaluations reveal reliable communication (< 10\% BER) and high SNR gain (< 40dB) under various practical settings with abundant ambient noise and obstruction. Trace-driven evaluation demonstrates concurrent communication of 1100 tags while keeping BER approximately 1.5\% for the entire tags, showing substantial potential towards massive connectivity.
% %-------------------------------------------------------------------------------

% This paper presents \ours, a practical mmWave backscatter system. With an extreme sensitivity of -115 dBm, \ours is perform reliably under various practical settings with abundant ambient reflections and blockages. Coordination-free FDMA effortlessly scales to thousands of concurrent tags for further practicality. \ours reader was implemented with commodity radars and the tags were prototyped on PCB. The trace-driven evaluation demonstrates concurrent communication of 1100 tags at BER of $\sim$1.5\%.

\clearpage 

\balance
\bibliographystyle{abbrv}
\bibliography{ref}  % survey.bib is the name of the Bibliography in this case
\end{document}